\documentclass[reprint,aps,pra,twocolumn,superscriptaddress,nofootinbib]{revtex4-2}
\pdfoutput=1
\usepackage{lipsum} 
\usepackage{amsthm}
\usepackage{amssymb}
\usepackage{graphicx}

\usepackage[utf8]{inputenc}
\usepackage{amsfonts}

\usepackage{footmisc}
\usepackage{bbold}

\usepackage{mathrsfs}
\usepackage{verbatim}

\usepackage[caption=false]{subfig}

\usepackage[dvipsnames]{xcolor}
\usepackage{ragged2e}
\usepackage{halloweenmath} 
\usepackage{curve2e} 
\usepackage{hyperref}
\usepackage{bm}
\usepackage{float}

\newcommand{\ad}{\hat{a}^{\dagger}}

\newcommand{\ket}[1]{\left| {#1}\right\rangle}
\newcommand{\bra}[1]{\left\langle {#1}\right|}
\newcommand{\braket}[2]{\left\langle {#1}\middle|{#2} \right\rangle}

\newcommand{\ketbra}[2]{\left|{#1} \rangle  \langle {#2}\right|}

\newcommand{\perm}[1]{\text{perm}\left({#1}\right)}

\DeclareMathOperator{\Perm}{perm}
\newcommand{\derivate}[2]{\frac{\partial{#1}}{\partial{#2}}}

\newtheorem{conjecturep}{Conjecture~P\ignorespaces}
\newtheorem{conjecturem}{Conjecture~M\ignorespaces}

\begin{document}

\title{Limits of multimode bunching for boson sampling validation:\\  anomalous bunching induced by time delays}

\author{Léo Pioge}
\affiliation{Centre for Quantum Information and Communication, \'Ecole polytechnique de Bruxelles, CP 165/59, Universit\'e libre de Bruxelles, 1050 Brussels, Belgium}

\author{Leonardo Novo}
\affiliation{International Iberian Nanotechnology Laboratory (INL), Av. Mestre José Veiga, 4715-330 Braga, Portugal}

\author{Nicolas J. Cerf}
\affiliation{Centre for Quantum Information and Communication, \'Ecole polytechnique de Bruxelles, CP 165/59, Universit\'e libre de Bruxelles, 1050 Brussels, Belgium}

\begin{abstract}

The multimode bunching probability is expected to provide a useful criterion for validating boson sampling experiments. Its applicability, however, is challenged by the existence of anomalous bunching, namely paradoxical situations in which partially distinguishable particles exhibit a higher bunching probability in two or more modes than perfectly indistinguishable ones. Using multimode bunching as a reliable criterion of genuine indistinguishability, therefore, requires a clear identification of the interferometric configurations in which anomalous bunching can or cannot occur. In particular, since uncontrolled small time delays between single-photon pulses constitute a common source of mode mismatch in current photonic platforms, it is essential to determine whether the resulting photon distinguishability might lead to anomalous bunching. Here, we first identify a broad class of interferometric configurations in which anomalous bunching is rigorously excluded, thereby establishing regimes where multimode bunching-based validation remains valid. Then, we find that, quite unexpectedly, temporal mode mismatch does not belong to this class. We exhibit a specific interferometric setup in which temporal distinguishability enhances multimode bunching, demonstrating that time delays can induce an anomalous behavior. These results help clarify the conditions under which multimode bunching remains a reliable validation tool.

\end{abstract}

\maketitle
\vspace{-5pt}

\section{INTRODUCTION} \label{sec:Introduction}

Originally introduced by Aaronson and Arkhipov \cite{aaronson2011computational}, the boson sampling problem has emerged as a central milestone in the pursuit of a universal optical quantum computer. This computational task challenges the extended Church-Turing thesis, as it can be naturally implemented using linear optics while being widely believed to be intractable for classical computers. Several experimental implementations of boson sampling have been reported, demonstrating increasing levels of control and scalability \cite{Madsen:2022uqz,zhong2020quantum,Zhong_2021,deng2023gaussianbosonsamplingpseudophotonnumber,young2023atomic}. In realistic experimental conditions, implementations of boson sampling inevitably depart from the idealized setting due to various sources of noise, most importantly losses and particle distinguishability induced by imperfect mode matching. These imperfections can substantially reduce the computational complexity of the sampling task, potentially rendering it efficiently simulable on a classical computer \cite{PhysRevLett.120.220502,Go_2025,Qi_2020,hoven2025}. As a result, one of the central challenges in boson sampling experiments is to devise reliable validation methods capable of discriminating genuinely hard-to-simulate quantum data from data generated by noisy classical processes. Several validation strategies have been proposed to address this issue, including Bayesian hypothesis testing \cite{doi:10.1142/S021974991560028X,PhysRevA.103.023722,Flamini_2020}, cross-entropy \cite{PRXQuantum.5.040312,PhysRevLett.131.010401}, or detector binning \cite{Seron2024efficientvalidation,bressanini2024gaussianbosonsamplingvalidation,anguita2025experimentalvalidationbosonsampling}.


Among these, a very appealing approach relies on the intrinsic tendency of indistinguishable bosons to bunch \cite{shchesnovich2016universality}. Based on a combination of reasonable physical arguments \cite{shchesnovich2016universality,shchesnovich2016permanent} and mathematical results on matrix permanents \cite{bapat1985majorization}, one is led to expect that, given any linear interferometer, the probability for all bosons to be detected within a given subset of output modes is maximum when all bosons are fully indistinguishable (this is provably the case when bunching on a single output mode is considered). Measuring this multimode bunching probability should therefore hopefully provide a universal criterion to discriminate fully indistinguishable bosons from distinguishable ones.

However, contradicting such a physical intuition, specific instances of a linear interferometer have been found in Ref.~\cite{bosonbunching} such that well-chosen polarization-induced partial distinguishability leads to an enhancement of the two-mode bunching probability compared to the fully indistinguishable case. These interferometric configurations were obtained by building on the violation of a conjecture on matrix permanents \cite{drury2016}. This counterintuitive effect, which we refer to as anomalous boson bunching, implies that a multimode bunching-based validation protocol cannot be applied as such in a fully universal manner. Nevertheless, numerical investigations indicate that the generic behavior follows intuition \cite{shchesnovich2016universality} and that anomalous bunching only appears for highly specific interferometric configurations \cite{bosonbunching}, suggesting that this protocol may still be applicable to a broad class of experimentally relevant setups. Determining the precise conditions under which anomalous bunching can or cannot occur is therefore essential to assess the practical usefulness of the multimode bunching criterion.

Initial steps in this direction have already been taken, both by extending the class of states for which multimode bunching remains a faithful criterion \cite{geller2025} and by identifying other anomalous bunching configurations \cite{pioge2023anomalous}. In the former work, it was shown that, on average and under some assumptions, fully indistinguishable bosons maximize multimode bunching if one restricts to permutation-invariant states.  
In the latter work, it was shown that even nearly indistinguishable bosons—such as those typically encountered in experiments—can exhibit anomalous bunching. There, photon distinguishability was induced by polarization, which left open the question of whether other experimentally relevant sources of distinguishability, such as temporal or frequency mismatch, can also give rise to anomalous bunching.

In this work, we aim to better understand the boundary between interferometric configurations that provably obey intuitive boson bunching and those that allow for anomalous bunching. First, we significantly extend the set of known input states and linear interferometers for which this effect does not occur (see Sec.~\ref{Validation_of_Boson_Samplers}). Second, approaching the problem from the opposite perspective, we question whether anomalous bunching can occur when partial distinguishability arises solely from time delays (see Sec.~\ref{sec:Time_delay_distinguishability}).
Unexpectedly, we answer this question in the affirmative by exhibiting a 16-photon interferometric setup in which temporally-mismatched photons display a larger two-mode bunching probability than perfectly indistinguishable photons. Altogether, our results help clarify the situations in which multimode bunching provides a reliable validation tool and those in which it fails.


\section{BOSON BUNCHING IN LINEAR OPTICS}
\label{sec:Linear_interferometry}
\subsection{Preliminaries}
\begin{figure}[t]
    \centering
    \includegraphics[width = 0.49\textwidth]{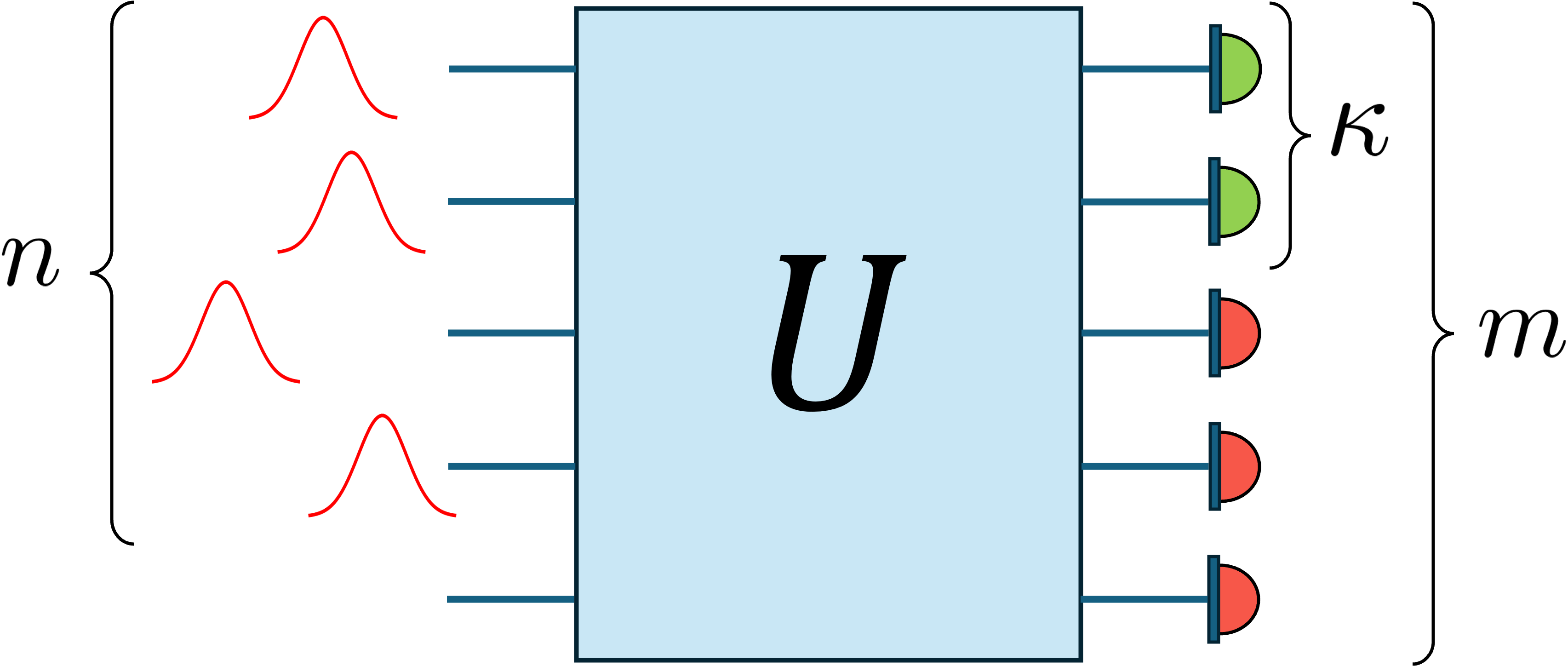}
    \caption{\justifying Quantum interferometric setup where $n$ single photons enter the first $n$ spatial modes of an $m$-mode linear interferometer $U$. The photon entering the $i$th mode carries specific internal degrees of freedom, represented by the internal state $\ket{\phi_i}$. The multimode bunching probability $P_{\kappa}$ corresponds to the probability that all $n$~photons are detected in the subset $\kappa$ of output modes (indicated by green detectors).} 
\label{fig:generalInterferometer}
\end{figure}

This section introduces the fundamental concepts of linear interferometry and boson bunching that will serve as the basis for the subsequent analysis. Consider a scenario where $n$ bosons enter a linear interferometer with $m$ modes, represented by $U$ (see Figure~\ref{fig:generalInterferometer}). We focus on the case in which a single photon is injected into each of the first $n$ input modes. The photons are in a pure product state. The creation operator associated to the state of the photon entering the $i$th input mode is denoted $a^{\dagger}_{i,\phi_{i}}$, such that  $a_{i,\phi_{i}}^{\dagger}\ket{0}=\ket{i}\otimes\ket{\phi_i}$. Here, $\ket{i}$ describes the spatial mode of the photon, while $\ket{\phi_i}$ denotes its internal state, encompassing all degrees of freedom unrelated to the spatial mode, such as polarization, time delay, or frequency. The interferometer $U$ acts on the spatial modes of the photons while leaving their internal states unchanged, namely (see Appendix~\ref{appendix:Derivation_of_the_Multimode_bunching_probability})
\begin{equation}\label{eq:assumptionU}
    \hat{U} \, \ad_{j, \phi} \, \hat{U}^{\dagger}= \sum_{k=1}^m U_{k,j} \, \ad_{k, \phi},~~\forall j, \forall \phi.
\end{equation}
In this work, we consider the (generalized) multimode bunching probability $P_\kappa$, defined as the probability that all photons are detected within a subset $\kappa$ of the output modes. According to Ref.~\cite{shchesnovich2016universality}, the multimode bunching probability can be expressed as (see Appendix~\ref{appendix:Derivation_of_the_Multimode_bunching_probability})
\begin{equation}
\label{eq:multimode_boson_bunching}
    P_\kappa= \perm{H\odot S}.
\end{equation}
Here,  $H$ is an $n \times n$ positive
semidefinite Hermitian ($p.s.d.h.$) matrix such that 
\begin{equation}
H_{i,j}=\sum_{k\in\kappa}U^{*}_{k,i}U_{k,j},~~\forall i,j.
\end{equation}
This matrix encapsulates all dependence on the interferometer and subset $\kappa$, whereas the matrix $S$, called the distinguishability matrix, is an $n\times n$ Gram matrix \cite{tichy2015_partial_distinguishability} that encapsulates the dependence on the internal states via their pairwise overlaps, 
 \begin{equation}
     S_{i,j}=\braket{\phi_i}{\phi_j},~~\forall i,j.
\end{equation}
Quantum states are defined up to local phases, $\ket{\phi_j} \rightarrow e^{i\theta_j}\ket{\phi_j}$, which does not affect event
probabilities, implying the existence of an equivalence class of distinguishability matrices corresponding to the same physical situation, related by gauge transformations $S_{i,j} \rightarrow e^{i(\theta_j - \theta_i)}S_{i,j}$. In Eq.~\eqref{eq:multimode_boson_bunching}, $\perm{\cdot}$ denotes the matrix permanent, and $\odot$ the Hadamard (element-wise) product, i.e., $(H\odot S)_{i,j}=H_{i,j}S_{i,j}$.  


\subsection{Anomalies in multimode bunching}

Since the partial distinguishability of photons is one of the main factors that can render boson sampling classically simulable, any boson sampling experiment should incorporate a validation criterion ensuring that the photons are sufficiently indistinguishable. At first glance, the measurement of the single-mode bunching probability appears to be a natural candidate. This event corresponds to the situation in which all photons exit through a single output mode of the linear interferometer. In this case, Eq.~\eqref{eq:multimode_boson_bunching} reduces to \cite{tichy2015_partial_distinguishability}
\begin{equation}
\label{eq:single-mode_boson_bunching}
   P_{k}=P^{d}_k~\perm{S},
\end{equation}
where $P_k$ denotes the single-mode bunching probability in mode $k$ and $P_k^{d}$ denotes the same for distinguishable particles. The factor $\perm{S}$ provides a well-known quantitative measure of indistinguishability \cite{shchesnovich2015tight, tichy2015_partial_distinguishability, shchesnovich2016universality}; in particular, $\perm{S}/n!$ corresponds to the weight of the permutationally symmetric component of the state $\ket{\Phi}= \ket{\phi_1}\ket{\phi_2}\dots\ket{\phi_n}$. Measuring the single-mode bunching probability, therefore, gives direct access to the degree of photon indistinguishability and can, in principle, be used as a validation test for boson sampling. Unfortunately, this approach is not practical since all $P^{d}_k$ decay exponentially as the number of photons $n$ increases \cite{arkhipov2011bosonicbirthdayparadox}.
A possible way out could be to turn to multimode bunching as the probability does not decay so fast with $n$, provided the subset $\kappa$ may include an arbitrary number of output modes, see Ref.~\cite{shchesnovich2016universality}. However, in this more general setting, the connection between photon indistinguishability and multimode bunching is no longer straightforward, and, at this stage, it can only be formulated as a conjecture.

\begin{conjecturep}[Multimode bunching conjecture]
\label{conjp:generalizedBunching}
Consider any linear interferometer $U$ and any (nontrivial) subset $\kappa$ of output modes. Among all possible separable input states of $n$ classically correlated photons, the probability that all output photons bunch into subset $\kappa$ reaches its \emph{global} maximum if the photons are perfectly indistinguishable. 
\end{conjecturep} 
This conjecture is supported by a strong physical intuition. Indeed, the enhancement of boson bunching events originates from quantum interference effects, which are maximized when the particles are fully indistinguishable. Refs.~\cite{shchesnovich2016universality,shchesnovich2016permanent} provide compelling evidence supporting the validity of the following Conjecture~P\ref{conjp:generalizedBunching}. This conjecture can be reformulated as the following mathematical statement: for any $H$ and $S$, we have
\begin{equation}
\label{eq:conjectureP1}
   \perm{H \odot S} \leq \perm{H}.
\end{equation}
When all photons are perfectly indistinguishable, the matrix $S$ reduces to $\mathbb{E}$ (where $\mathbb{E}_{i,j} = 1$, $\forall i,j$), so that $\perm{H \odot \mathbb{E}} = \perm{H}$. Beyond its intuitive physical interpretation, this conjecture was shown in  Ref.~\cite{shchesnovich2016universality} to be mathematically equivalent to a long-standing open problem formulated by R. B. Bapat and V. S. Sunder, who introduced a permanental analogue of the Oppenheim inequality \cite{oppenheim1930inequalities}.

\begin{conjecturem}[Bapat-Sunder 1985 \cite{bapat1985majorization}]
\label{conjm:bapatSunder_1}
Let $A$ and $B$ be $n\times n$ positive semidefinite Hermitian matrices, then  
\begin{equation}
\label{eq:conj1}
    \Perm\,(A\odot B) \leq \Perm\,(A) \prod_{i=1}^n B_{ii}.
\end{equation}
\end{conjecturem} 

In 2016, over three decades after its proposal, S. Drury provided a $7 \times 7$ counterexample to the Bapat-Sunder conjecture \cite{drury2016}. This mathematical counterexample was later used in Ref.~\cite{bosonbunching} to construct a family of physical counterexamples to Conjecture~P\ref{conjp:generalizedBunching}. The authors identified counterintuitive interferometric scenarios with seven or more photons, where partial distinguishability results in a higher bunching probability than full indistinguishability, a phenomenon which we refer to as anomalous bunching (see also  Ref.~\cite{pioge2023anomalous}).

A closer analysis of the indistinguishability measure based on $\perm{S}$ in the counterexamples found in Ref.~\cite{bosonbunching} reveals that the states exhibiting anomalous bunching deviate substantially from fully indistinguishable bosons. This observation, supported by additional arguments, led the authors of Ref.~\cite{pioge2023anomalous} to conjecture that the state of indistinguishable bosons still plays a privileged role in bunching phenomena. Specifically, while not being a global maximum, this state might nevertheless constitute a local maximum of the bunching probability, which can be expressed as a weaker (than P\ref{conjp:generalizedBunching}) conjecture.
\begin{conjecturep}[Local multimode bunching conjecture \cite{pioge2023anomalous}] \label{conjp:physical_conj}
Consider any linear interferometer $U$ and any (nontrivial) subset $\kappa$ of output modes. Starting from the product state of $n$ indistinguishable photons, the probability that all output photons bunch into a subset $\kappa$ can only decrease if an independent infinitesimal perturbation is applied to the internal state of each photon, making them slightly distinguishable. 
\end{conjecturep}

Since boson sampling experiments aim to prepare photons as indistinguishable as possible, the validity of this conjecture would allow multimode bunching-based validation protocols to be applicable in most practical cases. In Ref.~\cite{pioge2023anomalous}, the authors further demonstrated that Conjecture~P\ref{conjp:physical_conj} implies another permanental conjecture, also due to R. B. Bapat and V. S. Sunder.
\begin{conjecturem}[Bapat-Sunder 1986 \cite{bapat1986extremal}]
\label{conjm:bapatSunder_2}
Let $A$ be a positive semidefinite Hermitian matrix and let $F^A$ be the $n\times n$ matrix
defined by $F^A_{i,j}=A_{i,j} \, \Perm\,(A(i;j))$, then  
\begin{equation}
\label{eq:conjm_BS2}
\lambda_{\max}(F^{A}) \leq \Perm\,(A).
\end{equation}
\end{conjecturem}
Here, $\lambda_{\max}(F^{A})$ denotes the maximal eigenvalue of $F^{A}$ and $A(i; j)$ stand for the $(n - 1)\times (n - 1)$ submatrix of $A$ obtained by deleting the $i$th row and $j$th column of $A$. The authors of Ref.~\cite{pioge2023anomalous} then found a counterexample to Conjecture~P\ref{conjp:physical_conj}, consisting of an eight-photon configuration where nearly indistinguishable photons bunch more than fully indistinguishable ones. This construction was based on a counterexample to Conjecture~M\ref{conjm:bapatSunder_2}, discovered by S. Drury in Ref.~\cite{drury2018}.
Thus, the existence of anomalous bunching phenomena challenges the use of multimode bunching  as a tool to validate photon indistinguishability even in the practically relevant regime of near indistinguishability.

\section{Multimode bunching as a validation criterion}
\label{Validation_of_Boson_Samplers}

The main limitation of using multimode bunching as a validation criterion for any boson sampling experiment stems from its reliance on the validity of the multimode bunching conjecture~P\ref{conjp:generalizedBunching}, which is known to be false in full generality, i.e., for arbitrary interferometers $U$, arbitrary subsets $\kappa$ of output modes, and arbitrary internal photon states (distinguishability matrix $S$). Nevertheless, numerical simulations suggest that Conjecture~P\ref{conjp:generalizedBunching} remains valid for a broad class of interferometric configurations. Identifying these subclasses would enable the use of multimode bunching as a reliable validation tool within these restricted regimes. In this section, we first review the known scenarios in which multimode bunching reflects the indistinguishability of the particles, and then extend these results by identifying broader classes of configurations that prevent anomalous bunching.

\subsection{Condition on the interferometer $U$ and subset $\kappa$}

We seek conditions on the interferometer $U$ and subset $\kappa$  of output modes that ensure that multimode boson bunching reflects particle indistinguishability for arbitrary internal photon states. In other words, we are interested in the known conditions on $H$ such that 
\begin{equation}
    \perm{H\odot S}\leq \perm{H},
\label{eq:condition-on-H-matrix}
\end{equation} 
holds for any distinguishability matrix $S$. 

\subsubsection{Rank-one $H$ matrices}
If $H$ is a rank-one matrix, the Bapat-Sunder inequality holds for any distinguishability matrix $S$ \cite{zhang2013,Math_paper}. In this case, $H$ can be written as $H_{i,j}=h_i^{*} h_j$ for all $i,j$. As a consequence, all elements of row $i$ of $H$ share the common factor $h_i^{*}$, while all elements of column $j$ of $H$ share the common factor $h_j$, all these factors can be factored out of the permanent. By applying the same argument successively to all rows and columns, one obtains
\begin{align}
\perm{H\odot S}
&=\prod_{i=1}^n H_{i,i}\perm{\mathbb{E}\odot S} \nonumber\\
&=P^{d}_{\kappa}~\perm{S}
\end{align}
Here, the prefactor $P^{d}_{\kappa}=\prod_{i=1}^n H_{i,i}$ corresponds to the bunching probability of distinguishable particles. When $H$ has rank one, the bunching probability of the photons behaves as $\perm{S}$, which quantifies their degree of distinguishability. This observation naturally accounts for the behavior of single-mode bunching \cite{tichy2015_partial_distinguishability}, since the matrix $H$ associated with single-mode bunching into a spatial mode $k$ can be written as $H_{i,j}=U_{k,j}^{*}U_{k,i}$. More generally, the mathematical constraint $\mathrm{rank}(H)=1$ can also be realized in multimode bunching scenarios. It suffices to choose an interferometer $U$ such that all matrix elements $U_{k,i}$ depend either solely on $k$ or solely on $i$, for all $k\in \kappa$ and $i\in \{1,\dots,n\}$. An example of such a configuration is provided by cascading two interferometers. First, $n$ photons are injected into a $m_1$-mode interferometer $U_1$. One output mode is then split into $m_2$ modes by using a second interferometer $U_2$, with the remaining $m_2-1$ input modes of $U_2$ being in the vacuum. 
The overall circuit, therefore, can be viewed as a single interferometer acting on $m_1 + m_2 - 1$ modes. In this cascaded configuration, considering the multimode bunching probability that all $n$ photons are detected within a subset $\kappa$ consisting of the $m_2$ output modes of $U_2$ leads to a rank-one matrix $H$, hence Eq.~\eqref{eq:condition-on-H-matrix} holds.

\subsubsection{nonnegative $H$ matrices and their equivalence class}
\label{sec:validity_H_pos}
In Ref.~\cite{zhang1989notes}, F. Zhang proved that if $H$ is a nonnegative matrix, i.e., $H_{i,j}\ge 0$ for all $i,j$, then inequality~\eqref{eq:condition-on-H-matrix}
holds for any distinguishability matrix $S$. Similarly to $S$, the matrix $H$ is also defined up to a gauge choice. Indeed, phase shifters can be added at each input port of an interferometer $U$ without affecting any physical probability. Introducing a phase $\theta_i$ at input mode $i$ results in the multiplication of all elements of column $i$ of $U$ by the phase factor $e^{i\theta_i}$. Consequently, two matrices $H$ related by the phase transformation
\begin{equation}
    H_{i,j}\rightarrow e^{i(\theta_j-\theta_i)}H_{i,j}=\Theta_{i,j}H_{i,j},~~\forall i,j, 
\end{equation}
belong to the same equivalence class. Here, $\Theta$ denotes the $n\times n$ rank-one matrix defined by $\Theta_{i,j}=e^{i(\theta_{j} -\theta_{i})}$ for all $i,j$. Zhang's condition can therefore be extended to any matrix $H$ belonging to the equivalence class of a nonnegative matrix. It is thus important to determine whether a given matrix $H$ lies in this class; to this end, we write
\begin{equation}
    H=R\odot \Phi,
\end{equation}
where $R_{i,j} = |H_{i,j}|$ and $\Phi_{i,j} = e^{i \arg(H_{i,j})}$. The matrix $H$ belongs to the equivalence class if and only if there exists a phase matrix $\Theta$ such that $\Theta\odot\Phi=\mathbb{E}$. Since $\Theta$ is rank one, this condition holds if and only if $\mathrm{rank}(\Phi) = 1$. Characterizing the class of unitaries and subsets $\kappa$ for which this condition is satisfied is nontrivial. However, it is straightforward to see that the condition is fulfilled whenever $ U_{i,j}\geq 0, \forall i \in \kappa,$ and $\forall j \in \{1,\dots,n\}$.

Incidentally, it is natural to ask whether inequality~\eqref{eq:condition-on-H-matrix} may also hold when $S$ belongs to the equivalence class of nonnegative matrices, instead of $H$. This would have consequences for multimode bunching when photon distinguishability is induced by temporal mismatch (time delays). We shall answer this question in the negative, see Section~\ref{sec:Time_delay_distinguishability}.

\subsection{Conditions on photon distinguishability}

After investigating the restrictions on $H$ (hence, on $U$ and $\kappa$), we now investigate conditions on the photon states that rule out anomalous bunching for any interferometer and any subset of output modes. In particular, we present several photon distinguishability models that yield a distinguishability matrix $S$ satisfying
\begin{equation}
\perm{H\odot S}\leq \perm{H},
\label{eq:condition-on-S-matrix}
\end{equation}
for any matrix $H$. 
\subsubsection{One, two, or three interfering photons}
\label{sec:3_photons}
In Ref.~\cite{bapat1986extremal}, R. B. Bapat and V. S. Sunder proved that their inequality holds for matrices of dimension three or fewer. This result implies that anomalous bunching cannot occur for interferences involving up to three photons. Conversely, explicit counterexamples have been found with seven or more photons \cite{bosonbunching}, establishing the general invalidity of the conjecture. However, the status of the inequality \eqref{eq:condition-on-S-matrix}
for four, five, and six photons remains unresolved, despite its importance for present-day experimental implementations of anomalous bunching.

\subsubsection{The $x$-model of distinguishability}
The $x$-model provides a simple and widely used framework for modeling particle distinguishability \cite{PhysRevLett.120.220502,renema2019classical,Moylett_2020,Go_2025,robbio2024centrall}. This model corresponds to uniform pairwise distinguishability, interpolating between fully indistinguishable and fully distinguishable particles. More precisely, all photons initially share the same internal state $\ket{\chi}$ and are continuously interpolated towards the set of fully distinguishable states $\{\ket{\eta_1},\ldots,\ket{\eta_n}\}$,
\begin{equation}
     \ket{\phi_i}=x\ket{\chi}+\sqrt{1-x^2} \ket{\eta_i},~~\forall i.
\end{equation}
with $\braket{\eta_i}{\eta_j}=\delta_{i,j}$ and $\braket{\chi}{\eta_i}=0,~\forall i,j$. The parameter $x$ can be restricted to the real interval $[0,1]$, as any phase can be absorbed into the definition of $\ket{\chi}$. This model yields the following distinguishability matrix:
\begin{equation}
\label{eq:S_x_model}
    S^{x}_{i,j}=x^{2}(1-\delta_{i,j})+\delta_{i,j}.
\end{equation}
In Ref.~\cite{zhang1989notes}, F. Zhang showed that, for any matrix $H$, the value of $\perm{H\odot S^x}$ is no greater than $\perm{H}$. To do so, he proved that for all $x \in [0,1]$, the following inequality holds:
\begin{equation}
\label{eq:deriv_pos_xmodel}
    \derivate{\perm{H\odot S^x}}{x}\geq0.
\end{equation}
This result implies that if $x_1$ and $x_2$ are two values such that $0 \leq x_1 \le x_2 \leq 1$, then for any distinguishability matrix $H$, we have
\begin{align}
    \perm{H\odot S^{x_1}} &\le \perm{H\odot S^{x_2}} \nonumber\\ 
    &\leq \perm{H}.
\end{align}
This implies that, within the $x$-model, increasing indistinguishability (larger $x$) increases the bunching probability. The maximum is reached at $x=1$, corresponding to fully indistinguishable particles.

\subsubsection{The $x_i$-model of distinguishability}
\label{sec:x_i-model}
In the $x$-model, the interpolation goes from the starting point of indistinguishable photons toward the point of fully distinguishable photons using the same coefficient $x$ for each photon. This restriction is a significant limitation for modeling partial distinguishability. Here, we broaden the class of states that satisfy inequality \eqref{eq:condition-on-S-matrix} by proving that removing this constraint, i.e., allowing photon-dependent interpolation coefficients, still preserves the inequality. This physical configuration is described by the following quantum states and the associated distinguishability matrix:
\begin{align}
\label{eq:S_more_general_interpolation}
    \ket{\phi_i}&=x_i\ket{\chi}+\sqrt{1-|x_i|^2} \ket{\eta_i}\nonumber \\
    S^{\bm{x}}_{i,j}&=x_i^{*}x_j(1-\delta_{i,j})+\delta_{i,j}
\end{align}

This model has been used to describe photon distinguishability in Ref.~\cite{hoven2025}. 
As before, since the phase of each coefficient $x_i$ can be absorbed into the corresponding state $\ket{\eta_i}$, we may, without loss of generality, restrict to real $x_i \in [0,1]$ for all $i$. In Appendix~\ref{appendix:x_i-model}, we prove that:
\begin{equation}
\label{eq:deriv_pos_xi}
\derivate{\perm{H\odot S^{\bm{x}}}}{x_i}\geq 0,~ \forall i.
\end{equation}
Hence, the bunching probability is maximized when $x_i=1$ for all $i$, i.e., when all particles are fully indistinguishable.

\subsubsection{More general interpolation toward fully \\distinguishable bosons}
\label{final_interp_toward_dist}
In the $x_i$-model, the coefficients involved in the interpolation toward the point of fully distinguishable photons are different, but the starting point corresponds to fully indistinguishable photons $\ket{\chi}^{\otimes n}$, that is, all photons are in the same initial state.
Instead, we now examine the more general situation where the starting point is such that the photons are initially in the state $\{\ket{\chi_1},\ldots,\ket{\chi_n}\}$, with $\braket{\chi_i}{\chi_j}=S^{\chi}_{i,j}$. This initial configuration is assumed to satisfy the Bapat-Sunder inequality for any matrix $H$, i.e., $\perm{H\odot S^{\chi}}\leq \perm{H}$. As before, an interpolation is then performed toward fully distinguishable states $\{\ket{\eta_1}\ldots\ket{\eta_n}\}$, with $\braket{\eta_i}{\eta_j}=\delta_{i,j},~\forall i,j$. These states are also orthogonal to all initial states, i.e., $\braket{\eta_i}{\chi_i}=0, ~\forall i$. This results in the following configuration:
\begin{align}
\label{eq:xi_model}
    \ket{\phi_i}&=x_i\ket{\chi_i}+\sqrt{1-|x_i|^2} \ket{\eta_i}\nonumber \\
    S_{i,j}&=\braket{\chi_i}{\chi_j} \,x_i^{*}x_j(1-\delta_{i,j})+\delta_{i,j}.
\end{align}
Observe that $S=S^{\chi}\odot S^{\bm{x}}$, where $S^{\bm{x}}$ is defined in Eq.~\eqref{eq:xi_model}. Since we have already shown in Sec.~\ref{sec:x_i-model} that $S^{\bm{x}}$ satisfies the Bapat-Sunder conjecture for any $p.s.d.h.$ matrix, the result follows directly
\begin{align}
    \perm{H\odot S}&=\perm{(H\odot S^{\chi}) \odot S^{\bm{x}}} \nonumber \\
    &\leq \perm{H\odot S^{\chi}} \nonumber \\
    &\leq\perm{H}.
\end{align}

Note that, due to Eq.~\eqref{eq:deriv_pos_xi}, interpolating toward distinguishable particles can only decrease the bunching probability. Consequently, it is not even necessary for $S^{\chi}$ to satisfy Eq.~\eqref{eq:condition-on-S-matrix} for all matrices $H$. If $S^{\chi}$ satisfies the Bapat-Sunder inequality for a specific matrix $H$, then the state described by Eq.~\eqref{eq:xi_model} will never exhibit anomalous bunching for this particular $H$.

\subsubsection{Two non-orthogonal sets of fully indistinguishable bosons}
We now consider the case of $n$ photons grouped into two sets of fully indistinguishable particles, containing $k$ and $(n-k)$ photons, respectively, while the two sets are only partially distinguishable from one another. Specifically, the internal states are given by
\begin{equation}
\ket{\phi_i}=
 \begin{cases}
\ket{\chi}& \text{if}~ 1 \leq i\leq k \\
\ket{\eta}& \text{if}~ k+1 \leq i\leq n. 
\end{cases}
\end{equation}
Assuming $\braket{\eta}{\chi}=x$, the corresponding distinguishability matrix $S(x)$ takes a block form
\begin{equation}
\label{eq:block_matrix}
    S=\begin{pmatrix}
        \mathbb{E}^{k} & X \\
        X^{\dagger} & \mathbb{E}^{(n-k)}
    \end{pmatrix}.
\end{equation}
Here, $\mathbb{E}^{a}$ denotes the $a\times a$ all-ones block, and $X$ is the $k\times(n-k)$ block with all entries equal to $x$. It is shown in Appendix~\ref{appendix:2_set} that
\begin{equation}
\label{eq:deriv_pos}
\derivate{\perm{H\odot S(x)}}{x}\geq 0.
\end{equation}
As in previous cases, the bunching probability is maximized at $x=1$, i.e., for fully indistinguishable particles. 

At this stage, it is natural to ask whether the generalization to three non-orthogonal sets of fully indistinguishable particles $\{ \ket{\chi} \dots \ket{\chi},\ket{\eta}\dots \ket{\eta},\ket{\mu}, \dots \ket{\mu}\}$ may also satisfy inequality \eqref{eq:condition-on-S-matrix}. In full generality, this situation involves three independent overlaps, namely $\langle\chi|\eta\rangle = x$, $\langle\chi|\mu\rangle = y$, and $\langle\eta|\mu\rangle = z$. However, it remains unclear whether the generalization of Eq.~\eqref{eq:deriv_pos} holds in this case, leaving this scenario as an open question.

\subsubsection{$k$ orthogonal sets of bosons}
\label{sec:k_orthogonal_set_of_dist}

Let us consider the case of $k$ orthogonal sets of particles that do not exhibit anomalous bunching when considered separately, for instance, because they belong to the class of states introduced in the previous subsections. In this situation, the distinguishability matrix $S$ can be expressed as a direct sum of $k$ distinguishability matrices 
\begin{equation}
\label{eq:S_for_d_distinguishable_part}
    S= \bigoplus_{i=1}^k S^{(i)}.
\end{equation}
Here $S^{(i)}$ denotes the distinguishability matrix associated with the 
$i$th subset of particles. The Hadamard product between $H$ and $S$ then decomposes as
\begin{equation}
H\odot S = \bigoplus_{i=1}^k \left(H^{(i)}\odot S^{(i)}\right),
\end{equation}
where $H^{(i)}$ denotes the principal submatrix of $H$ associated with the $i$th subset of particles. By hypothesis, each matrix $S^{(i)}$ satisfies the Bapat-Sunder inequality for any matrix $H$. We therefore obtain
\begin{align}
    \perm{H\odot S}&=\prod_{i=1}^k\perm{H^{(i)}\odot S^{(i)}}\nonumber\\
    &\leq\prod_{i=1}^k\perm{H^{(i)}} \nonumber\\
    &\leq\perm{H}.
\end{align}
The first inequality follows from the Bapat-Sunder inequality, while the last step relies on Lieb’s inequality~\cite{lieb1966}. Thus, the distinguishability matrix \eqref{eq:S_for_d_distinguishable_part} cannot give rise to anomalous bunching for any matrix $H$.

\subsubsection{Interpolation toward orthogonal sets of fully indistinguishable particles}

As in the previous configuration, it is possible to significantly extend the class of states satisfying the Bapat-Sunder inequality by concatenating several matrices that individually satisfy the conjecture. Concatenating $k$   matrices following  Eq.~\ref{eq:block_matrix}, with different subset sizes, results in an $n \times n$ distinguishability matrix $S$, partitioned into $k\times k$ blocks $S^{(i,j)}$ for $i,j=1,\dots,k$. Each block $S^{(i,j)}$ has dimension $k_i\times k_j$, with $\sum_{i=1}^k k_i=n$, and constant entries given by $x_i^{*}x_j(1-\delta_{i,j})+\delta_{i,j}$. Such a distinguishability matrix physically corresponds to $k$ sets of particles, where the $k_i$ particles within set $i$ all share the same internal state $\ket{\phi_i}=x_i\ket{\chi}+\sqrt{1-|x_i|^2}\ket{\eta_i}$, with $\braket{\eta_i}{\eta_j}=\delta_{i,j}$ and $\braket{\chi}{\eta_i}=0$, for all $i,j$. As before, one may also allow the states $\ket{\chi}$ to differ for each particle, provided that the corresponding overlaps yield a $n\times n$ distinguishability matrix  $S^{\chi}_{a,b}=\braket{\chi_a}{\chi_b}$ satisfying the Bapat-Sunder inequality. This construction constitutes a further generalization of the model presented in Sec.~\ref{final_interp_toward_dist}, where identical $\ket{\eta_i}$ are allowed only under the condition of sharing the same distinguishability factor $x_i$. A natural next step toward extending the domain of validity of the conjecture would be to relax this last restriction. However, the counterexamples reported in Refs.~\cite{drury2016,bosonbunching,pioge2023anomalous} already arise within this specific configuration.

So far, we have shown that several models of partial distinguishability exhibit a lower bunching probability than the fully indistinguishable state. However, we emphasize that our results are stronger than this observation alone would suggest. Indeed, for all these models, we have proven that the bunching probability is a monotonic function of the indistinguishability parameter, as established in Eqs.~\eqref{eq:deriv_pos_xmodel}, \eqref{eq:deriv_pos_xi}, and \eqref{eq:deriv_pos}. This implies that these classes of states obey the intuitive rule that increasing the particle indistinguishability leads to a stronger tendency for particles to bunch together.

\subsubsection{Uniform mixed states}

Let us briefly leave the pure-state setting and turn to the case of uniform mixed states. This corresponds to a configuration in which the internal density matrix of each boson is identical, so that the internal density matrix of the full system is given by $\Omega=\rho^{\otimes n}$. This scenario is particularly relevant as it reflects the situation in which all bosons are produced by the same noisy source.

In Ref.~\cite{geller2025}, the authors showed that, on average, such states $\rho^{\otimes n}$ exhibit a lower bunching probability than fully indistinguishable bosons when the interferometer is sampled from the Haar measure. Although this result only provides a bound on the \textit{average} bunching probability, it was also proven in Ref.~\cite{geller2025} that this bound holds for any interferometer (for any $H$) if Lieb’s permanental-dominance conjecture for immanants holds \cite{lieb1966}. This is because states $\rho^{\otimes n}$ are a special case of the permutation-invariant states (as defined in Definition 3.3 of Ref.~\cite{geller2025}), which were proven to bunch less than fully indistinguishable bosons for any interferometric configuration by taking Lieb's conjecture for granted.

Here, we prove that uniform mixed states $\rho^{\otimes n}$ have a lower bunching probability than indistinguishable bosons for any interferometer and any choice of output-mode subset, without relying on Lieb’s conjecture. Without loss of generality, the internal density matrix of each photon can be written as $\rho=\sum_{k=1}^L{\alpha_k}|\lambda_k \rangle \! \langle \lambda_k|$, where $L$ denotes the total number of internal degrees of freedom and $\{\ket{\lambda_k}\}_{k=1}^{L}$ is an orthonormal basis, yielding $\sum_{k}\alpha_{k}=1$ and $\alpha_{k}\geq 0$ for all $k$. The total internal density matrix $\Omega$ can then be expressed as
\begin{equation}
\Omega=\sum_{\bm{j}}\alpha_{\bm{j}}|\lambda_{j_1} \cdots \lambda_{j_n} \rangle \! \langle \lambda_{j_1} \cdots \lambda_{j_n}|,
\end{equation}
where $\alpha_{\bm{j}}=\prod_{i=1}^{n}\alpha_{j_i}$.
According to Ref.~\cite{shchesnovich2016universality}, the generalization of Eq.~\eqref{eq:multimode_boson_bunching} to mixed states is
\begin{equation}
\label{eq:bunching_mixed_states}
    P_{\kappa}= \sum_{\bm{j}}\alpha_{\bm{j}}\perm{H\odot S^{\bm{j}}},
\end{equation}
where $S_{a,b}^{\bm{j}}= \braket{j_a}{j_b}=\delta_{j_a,j_b}$. Owing to this structure,

there exists a permutation matrix $P_\sigma$, such that the matrix $\tilde{S}^{\bm{j}}=P_{\sigma}S^{\bm{j}}P_{\sigma}^{T}$ belongs to the class of matrices described in Eq.~\eqref{eq:S_for_d_distinguishable_part}, which was shown to satisfy the Bapat-Sunder inequality.
Since the permanent is invariant under permutations of the rows and columns, it follows that $\perm{H\odot S^{\bm{j}}}=\perm{P_{\sigma}H P_{\sigma}^{T}\odot \tilde{S}^{\bm{j}}}$. Consequently,
\begin{align}
    P_{\kappa}&= \sum_{\bm{j}}\alpha_{\bm{j}}\perm{H\odot S^{\bm{j}}} \nonumber\\
    &=\sum_{\bm{j}}\alpha_{\bm{j}}\perm{P_{\sigma}H P_{\sigma}^{T}\odot \tilde{S}^{\bm{j}}} \nonumber    \\
    &\leq 
    \sum_{\bm{j}}\alpha_{\bm{j}}\perm{P_\sigma H P_\sigma^{T}} \nonumber\\
    &\leq\perm{H}.
\end{align}
Uniform mixed states, therefore, bunch less than indistinguishable particles for any interferometer $U$ and any subset $\kappa$ of output modes.

\subsubsection{Beyond the uniform mixed-state assumption}
We have shown that uniform mixed states exhibit less bunching than indistinguishable particles. However, our analysis is not restricted to uniform mixed states. In this section, we present four simple examples of non-uniform mixed states that also exhibit less bunching than in the indistinguishable case. In full generality, the internal density matrix of the system, $\Omega=\bigotimes_{i=1}^{n} \rho^{(i)}$, 
can always be expressed as a convex combination of pure product states,
\begin{equation}
\Omega=\sum_{\bm{j}}\alpha_{\bm{j}}\ketbra{\phi_{j_1}\dots \phi_{j_n}}{\phi_{j_1} \dots \phi_{j_n}}.
\end{equation}
The corresponding bunching probability takes the same form as in Eq.~\eqref{eq:bunching_mixed_states}, except that $S_{a,b}^{\bm{j}}$, is now an arbitrary Gram matrix defined by $S_{a,b}^{\bm{j}}=\braket{\phi_{j_{a}}}{\phi_{j_b}}$ \cite{shchesnovich2016universality}.
As before, the central argument of this section is that, owing to the structure of Eq.~\eqref{eq:bunching_mixed_states}, any mixed state that can be expressed as a convex mixture of pure input states whose Gram matrices satisfy the Bapat-Sunder inequality necessarily exhibits less bunching than indistinguishable bosons. The first scenario follows directly from this argument. It corresponds to mixed states composed of one, two, or three particles, which, according to the results of Section~\ref{sec:3_photons}, can never exhibit anomalous bunching.

The second scenario corresponds to the trivial case of fully distinguishable mixed states, i.e.,
$\braket{\phi_{j_{a}}}{\phi_{j_{b}}}=\delta_{j_a,j_b}$, for any $\bm{j}$. Here, all the distinguishability matrices $S^{\bm{j}}$ reduce to the identity matrix $\mathbb{1}$, which yields
\begin{align}
    P_\kappa=&\sum_{\bm{j}}\alpha_{\bm{j}}\perm{H\odot \mathbb{1}}\nonumber\\
    = &\sum_{\bm{j}}\alpha_{\bm{j}}\prod_{i}H_{i,i} \nonumber\\
    \leq& \perm{H},
\end{align}
where the last step follows from the Marcus inequality \cite{Marcus1963ThePA}.

The third scenario is a natural generalization of uniform mixed states. Here, we consider the case where all photons share the same orthonormal basis $\{\ket{\lambda_k}\}_{k=1}^{L}$, but have different eigenvalues, namely 
\begin{equation}
\rho^{(i)}=\sum_{k=1}^L\alpha_k^{(i)}\ketbra{\lambda_k}{\lambda_k}.
\end{equation}
We can therefore apply exactly the same reasoning as in the previous section to show that these states bunch less than indistinguishable bosons.

In the fourth scenario, we consider the case where each $\rho^{(i)}$ is a rank-two matrix of the form
\begin{equation}
\rho^{(i)}=\alpha_i\ketbra{\phi_1}{\phi_1}+(1-\alpha_i)\ketbra{\phi_2}{\phi_2},
\end{equation}
with $\braket{\phi_1}{\phi_2}=x$, which yields
\begin{equation}
    S_{a,b}^{\bm{j}}= \braket{\phi_{j_a}}{\phi_{j_b}}= x (1-\delta_{j_a,j_b})+\delta_{j_a,j_b}.
\end{equation}
Since these matrices belong, up to a permutation of rows and columns, to the class described in Eq.~\eqref{eq:block_matrix}, then each of them satisfies the Bapat-Sunder inequality.

\section{
TIME DELAYS AS A SOURCE OF ANOMALOUS BUNCHING
}
\label{sec:Time_delay_distinguishability}
In practice, photon distinguishability may arise from several physical sources, such as polarization, frequency, and time delays. All currently known anomalous bunching configurations involve a discrete two-dimensional internal degree of freedom (e.g., polarization) as the source of photon distinguishability \cite{bosonbunching,pioge2023anomalous}. However, uncontrolled time delays between single-photon pulses are a dominant source of mode mismatch in contemporary photonic platforms, which raises the experimentally relevant question of whether time delays can also give rise to anomalous bunching. This question is of particular interest for validation protocols. Indeed, if temporal distinguishability could only decrease bunching, then a validation protocol based on multimode bunching would provide a reliable means of certifying the correct behavior of a boson sampler affected by unwanted temporal-mode mismatch.

We begin in Section~\ref{sec:mathematical_conjecture} by identifying a mathematical condition on $p.s.d.h.$ matrices which, when fulfilled, implies the existence of anomalous bunching induced by time-delay distinguishability.  In Section~\ref{Enhancing_bunching_time_delays}, we exploit a  $16 \times 16$ matrix that fulfills this condition to propose an $18$-mode interferometer and a $16$-photon state for which an appropriate delay between the photons’ arrival times can surprisingly increase the two-mode bunching probability.

\subsection{Criterion for the existence of temporal anomalous bunching}
\label{sec:mathematical_conjecture}
A standard model for multiphoton interference assumes that each photon is described by a Gaussian wave packet centered at times $\{t_1,\ldots,t_n\}$, forming a real vector $\bm{t}$, where $\bm{\tau}=\bm{t}/||\bm{t}||$ denotes its normalized version. All wave packets are assumed to have identical variance $\sigma^2$. This model of distinguishability leads to the following distinguishability matrix \cite{MenssenDistinguishability}:
\begin{equation}
\label{eq:overlap_gaussian}
S^{\bm{\tau}}_{i,j}= e^{-(t_i-t_j)^2/(2\sigma)^2},~~\forall i,j.
\end{equation}
Within this framework, the occurrence of anomalous bunching in a physical system necessitates the presence of a pair of matrices, $H$ and $S^{\bm{\tau}}$, such that
\begin{equation}
\label{eq:BS1_with_time-delay}
    \perm{H\odot S^{\bm{\tau}}}> \perm{H}.
\end{equation}
The main particularity of this model is that up to a suitable choice of gauges, $S^{\bm{\tau}}_{i,j}\ge 0$ for all $i,j$, which precludes the counterexamples in Refs.~\cite{drury2016,bosonbunching,pioge2023anomalous,drury2017real} for implementation with time delays.
Furthermore, according to Section~\ref{sec:validity_H_pos}, if $H_{i,j}\ge 0$ for all $i,j$, it is known that for all $S^{\bm{\tau}}$ we have $\perm{ H \odot S}\leq \perm{H}$. 
Owing to the commutativity of the Hadamard product, one might hastily conclude that imposing the same positivity condition on $S$ instead of $H$ would likewise ensure the validity of the Bapat-Sunder inequality, thereby implying that Eq.~\eqref{eq:BS1_with_time-delay} can never be satisfied.

The goal of this section is to establish a mathematically necessary and sufficient condition for the occurrence of anomalous bunching due to time delay. While Eq.~\eqref{eq:BS1_with_time-delay} formally captures this condition, it remains impractical for direct application. To investigate this question, consider the delay strength $d=||\bm{t}||/(2\sigma)$, so $d=0$ corresponds to indistinguishable bosons. The distinguishability matrix then takes the form $S^{\bm{\tau}}_{i,j}=e^{-(\tau_i-\tau_j)^2 d^2}$. For convenience, define the $p.s.d.h.$  matrix $G(d)=H\odot S^{\bm{\tau}}(d)$. 
The existence of a matrix $G(d)$ such that 
\begin{equation}
\label{eq:derivate_sup_0}
    \derivate{\perm{G(d)}}{d}\geq 0,
\end{equation}
is a necessary condition for the occurrence of an anomalous bunching phenomenon with time delay. In what follows, we will demonstrate that, contrary to initial expectations, this condition is also sufficient. As shown in Appendix~\ref{appendix:Time_delay}, the derivative of the bunching probability with respect to the delay strength $d$ can be written as
\begin{equation}
\label{eq:derivate_time_delay}
\derivate{\perm{G(d)}}{d}= 4d\big(\bm{\tau}^T F^{G}\bm{\tau}-\perm{G(d)}\big),
\end{equation}
where $F^{G}$ is the $n\times n$ $p.s.d.h.$ matrix defined by $F_{i,j}^{G}=G_{i,j}\perm{G(i;j)}$, $\forall i,j$, with $G(i;j)$ denoting the $(n-1) \times (n-1)$ submatrix of $G$ obtained by deleting the $i$th row and $j$th column of $G$. Since $\bm{\tau}$ is a real vector, the vector $\bm{\tau}_{\max}$ which maximizes the derivative for a given interferometer and a fixed delay strength $d$, is the eigenvector associated with $\lambda_{\max}^{\mathbb{R}}(F^{G(d)})$, the largest eigenvalue of the symmetric part of $F^{G(d)}$ over $\mathbb{R}^{n\times n}$, i.e., the matrix formed by taking the real part of each entry of $F^{G}$. Since $d$ is a positive number, the necessary condition from Eq.~\eqref{eq:derivate_sup_0} can thus be restated as the simpler following statement: there exists a matrix $G(d)$, such that  
\begin{equation}
    \label{eq:condition_anomalous_bunching}
    \lambda_{\max}^{\mathbb{R}}(F^{G(d)}) > \perm{G(d)}.
\end{equation}

To prove that Eq.~\eqref{eq:condition_anomalous_bunching} is both a necessary and sufficient condition, we show that, starting from an interferometer $U$ and a time-delay matrix $S^{\bm{\tau}}(d)$ satisfying inequality~\eqref{eq:condition_anomalous_bunching}, one can construct a modified interferometer $\tilde U$ which, under suitable conditions, gives rise to anomalous bunching. Furthermore, we show that the resulting configuration recovers a scenario similar to Ref.~\cite{pioge2023anomalous}, namely an anomalous bunching effect occurring with nearly indistinguishable bosons. Consequently, this would also provide a counterexample to Conjectures~P\ref{conjp:physical_conj}.
Let us begin the proof by assuming the existence of a matrix $G(d)=H\odot S^{\bm{\tau}_{\max}}(d)$ satisfying Eq.~\eqref{eq:condition_anomalous_bunching}.
We then construct a new interferometer $\tilde{U}$, associated with a new spatial distinguishability matrix $\tilde{H}$ such that $\tilde{H}=G(d)$. Indistinguishable particles in this new interferometer would have a bunching probability of $\text{perm}(\tilde{H})$. Next, we apply a new time delays $S^{\bm{\tau}_{\max}}(\tilde{d})$ in the same direction $\bm{\tau}_{\max}$ as in the previous configuration. This yields a new total distinguishability matrix $\tilde{G}(\tilde{d})=\tilde{H}\odot S^{\bm{\tau}_{\max}}(\tilde{d})$, thus, at $\tilde{d}=0$ we have 
$\text{perm}(\tilde{G}(0))=\perm{G(d)}$ and 
$\lambda_{\max}^{\mathbb{R}}(F^{\tilde{G}(0)})=\lambda_{\max}^{\mathbb{R}}(F^{G(d)})$. The first derivative of the bunching probability of this new scenario satisfies
\begin{equation}
\label{eq:first_deriv=0}
    \left .\frac{\partial\text{perm}(\tilde{G}(\tilde{d}))}{\partial \tilde{d}} \right|_{\tilde{d}=0} = 0. 
    \end{equation}
However, according to Appendix~\ref{appendix:Time_delay} and by hypothesis, we have
    \begin{align}
    \label{eq:second_deriv>0}
    \left .\frac{\partial^2\perm{\tilde{G}(\tilde{d})}}{\partial \tilde{d}^2} \right|_{\tilde{d}=0}\!\!\!\!\!\!\!\!&=4\Big(\lambda_{\max}^{\mathbb{R}}(F^{\tilde{G}(0)})-\perm{\tilde{G}(0)}\Big)\nonumber\\ 
    &=4\Big(\lambda_{\max}^{\mathbb{R}}(F^{G(d)})-\perm{G(d)}\Big)\nonumber \\ 
    &> 0. 
\end{align}
Starting from indistinguishable bosons and applying a small time delay along $\bm{\tau}_{\max}$ results in an increase in the bunching probability. Indeed, at $\tilde{d}=0$ the first derivative being zero and the second derivative being positive, the bunching probability therefore grows for small time delays (i.e., $d\ll1$). This anomalous bunching configuration with time-delay distinguishability is therefore also a counterexample to Conjecture~P\ref{conjp:physical_conj}. This concludes the proof that Eq.~\eqref{eq:condition_anomalous_bunching} is both necessary and sufficient for time-delay-induced (or temporal) anomalous bunching. We thus have established a connection between temporal anomalous bunching and the following mathematical Conjecture~M\ref{conjm:BS2real}, which is itself a special case of Conjecture~M\ref{conjm:bapatSunder_2} that was originally formulated in Ref.~\cite{Math_paper}
\begin{conjecturem}
\label{conjm:BS2real}
Let $A$ be a positive semidefinite Hermitian matrix, then  
\begin{equation}
\label{eq:conjm3}
\lambda_{\max}^{\mathbb{R}}(F^{A}) \leq \Perm\,(A).
\end{equation}
\end{conjecturem}

At this point, it is interesting to note that finding counterexamples to this latter conjecture is numerically simpler than investigating Eq.~\eqref{eq:BS1_with_time-delay}, since Eq.~\eqref{eq:conjm3} involves only a single matrix.

\subsection{Enhancing bunching with small time delays}
\label{Enhancing_bunching_time_delays}

In Ref.~\cite{Math_paper}, the authors presented a numerically found matrix $A$ that provides a $16 \times 16$ counterexample to Conjecture~M\ref{conjm:BS2real}. This matrix is constructed as $A=M^{\dagger}M$, with $M=M_{R}+iM_{I}$, where

\begin{widetext}
\centering

\begin{align}
M_{R}&=\left(\begin{array}{ccccccccccccccccc}
 25 & -23 &  29 &  11 & -20 & 47 &  18 & 29 & 35 & -25 & -32 & -28 & -18 & 25 & 12 & -36 \\
  8 &  38 & -11 &  34 &  61 & 42 & -23 & 10 & 35 &  24 &  11 &   9 &  13 &   -9 & 34 &  22
\end{array}\right) \nonumber\\
M_I&=\left(\begin{array}{ccccccccccccccccc}
 30 &  20 &  51 & -43 & -11 & 47 &   4 & 27 &  -26 &  -2 &  11 &  37 &  64 & 26 &  -28 &  23 \\
   0 &  20 &  10 &   4 &  28 & 12 & -46 & 24 &  -43 &  10 & -17 & -63 & -23 & 50 &  -40 &  15
\end{array}\right). 
\label{eq:XR_XI}
\end{align}
 \end{widetext}
A numerical evaluation gives $\lambda_{\max}^{\mathbb{R}}(F^{A})  \approx 2.2632\times10^{64}$ and $\perm{A}\approx 2.1978 \times10^{64}$. These values yield a ratio $\lambda^{\mathbb{R}}_{\max}(F^A)/\perm{A}\approx 1.0298$. Using the connection established in the previous section, this counterexample can be used to construct a temporal anomalous bunching configuration. 

Let us turn now to the physical implementation of this counterexample. To determine the unitary, we follow the procedure outlined in Refs.~\cite{bosonbunching,pioge2023anomalous}.  The general idea is to embed $\sqrt{\gamma}M$ into the upper-left block of a unitary matrix $U$, where $\gamma$ is chosen such that $\lVert \gamma M^\dagger M \rVert=1$, yielding  $H=\gamma A$, with $\gamma\approx3.3767 \times10^{-5}$. Since $\mathrm{rank}(H)=2$, the corresponding physical configuration involves multimode bunching across two output modes; without loss of generality, we select the first two modes. To ensure that the rows of $U$ form an orthonormal set, $\sqrt{\gamma}M$ can be completed by adding two additional columns. This construction implies that the resulting unitary must be an $18 \times 18$ matrix. The remaining $16\times18$ block of $U$ is then completed with orthonormal rows orthogonal to the first two. Using the method introduced by Reck \textit{et al.} \cite{Reck1994}, this unitary can be implemented experimentally with $32$ beam splitters.

\begin{figure}[t]
    \centering
    \includegraphics[width = 0.5 \textwidth]{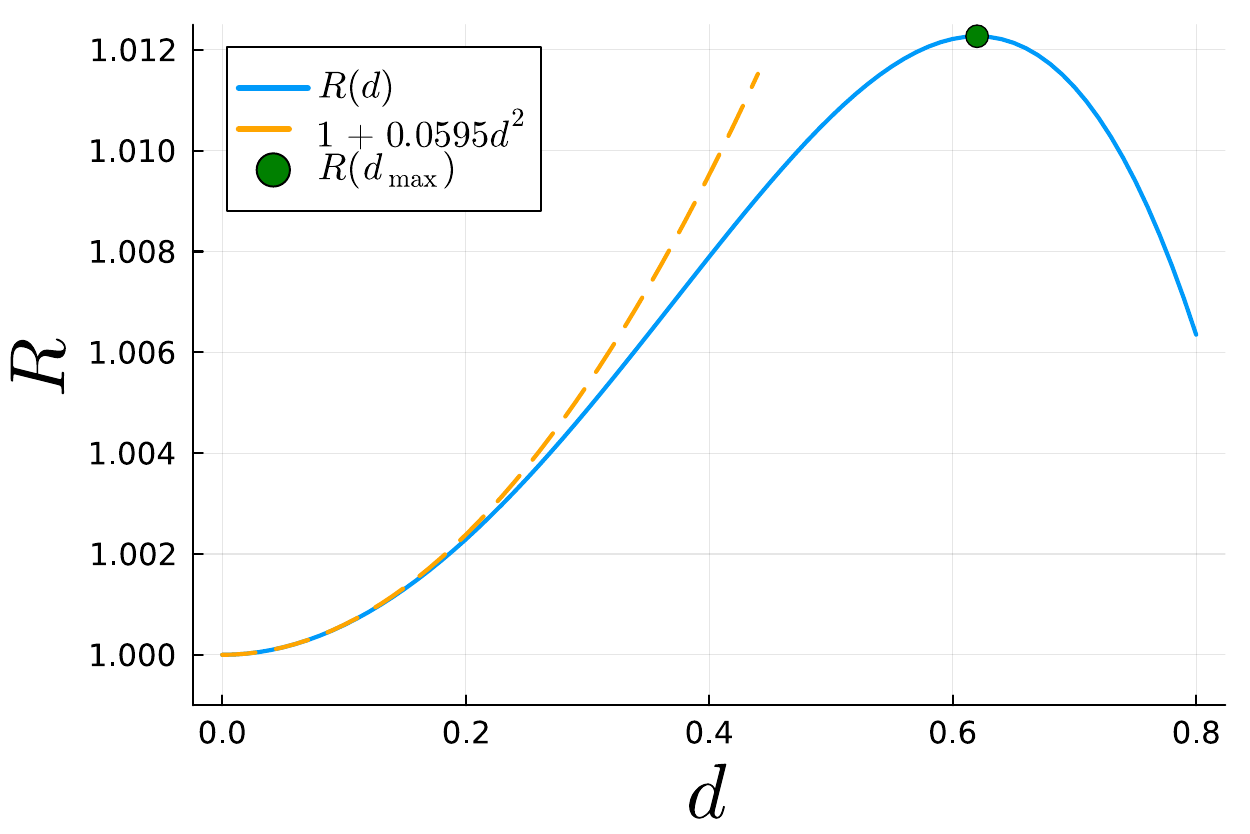}
    \caption{\justifying Violation ratio $R$ (solid line) as defined in Eq.~\eqref{eq:violation_ratio}, quantifying the relative enhancement of the bunching probability as a function of the perturbation strength of the delay $d$.  The perturbative approximation to second order $d^2$ is also plotted for comparison (dashed line). We observe that $R(d)$ exceeds $1$ in the neighborhood of indistinguishable particles, implying anomalous bunching.}
    \label{fig:ViolationRatio}
\end{figure}

The input state consists of $16$ photons prepared in Gaussian wave packets of identical variance $\sigma^2$, each centered at different times specified by the vector $\bm{t}(d)$, with $\bm{t}(d)=\frac{d}{2\sigma}\bm{\tau}_{\max}$, where $\bm{\tau}_{\max}$ is the eigenvector associated with the largest eigenvalue of the symmetric part of $F^{H}$ over $\mathbb{R}^{n\times n}$. From Eqs.~\eqref{eq:first_deriv=0} and \eqref{eq:second_deriv>0}, we obtain that the ratio between the two-mode bunching probability for slightly time-delayed photons and its counterpart for indistinguishable photons is given by
\begin{align}\label{eq:violation_ratio}
    R(d)&=\frac{ \perm{H\odot S^{\bm{\tau}_{\max}}(d)}}{\perm{H}}\nonumber \\
    &\approx  1+ 2 \left( \frac{\lambda_{\max}^{\mathbb{R}}(F^{H}) } {\perm{H}}-1\right) d^2 \nonumber \\
    & \approx 1+  0.0595 \,d^2. 
\end{align}
Since the ratio exceeds unity, the proof of the existence of anomalous bunching for small time delays is complete. Figure~\ref{fig:ViolationRatio} presents the numerical evaluation of the ratio $R(d)$.  The maximal value of the violation ratio $R(d_{\max})$ is approximately $1.0123$, corresponding to a violation of $1.23\%$, and is obtained for $d_{\max}=0.6201$. The corresponding absolute probabilities are $\perm{H}\approx 6.2797\times 10^{-8}$ and $\perm{H\odot(S^{\bm{\tau}_{\max}}(d_{\max}))}\approx 6.3568\times 10^{-8}$.

It is interesting to note that the same interferometer can also exhibit anomalous bunching when distinguishability arises from degrees of freedom other than time delays. Indeed, if the same distinguishability matrix as in Eq.~\eqref{eq:overlap_gaussian} is obtained through a different internal degree of freedom, the resulting violation ratio remains unchanged. This situation can be realized, for instance, by considering photons with Gaussian frequency distributions centered at different frequencies $\{\omega_1,\dots,\omega_n\}$ and sharing an identical variance $\sigma^2$. Furthermore, as $H$ is also a counterexample to Conjecture~M\ref{conjm:bapatSunder_2}, anomalous bunching can also be produced when polarization is used as the internal degree of freedom, by following the same procedure as in Ref.~\cite{pioge2023anomalous}.

In Ref.~\cite{Math_paper}, the authors report having performed numerical simulations to search for a counterexample to Conjecture~M\ref{conjm:BS2real} in lower dimensions. However, no such counterexample was identified, suggesting that the minimal number of photons required to observe this phenomenon may be sixteen.  Furthermore, in Ref.~\cite{drury2018}, S.~Drury stated that he was unable to find any counterexample to Conjecture~M\ref{conjm:bapatSunder_2} when restricted to real matrices, which led him to conjecture that the inequality holds in that case. This would imply that anomalous bunching induced by time delays is not possible in interferometers governed by orthogonal (real unitary) matrices, thereby allowing the use of multimode bunching as in Ref.~\cite{shchesnovich2016universality} to certify a boson sampler affected by time-delay distinguishability in this subclass of real unitaries.

\section{Conclusion}

\begingroup
\emergencystretch=2em
Motivated by the need for reliable validation protocols for boson sampling in the presence of realistic experimental imperfections, this work investigated the extent to which bunching probabilities can be used to discriminate genuinely indistinguishable bosons from partially distinguishable ones or from outcomes generated by noisy classical processes.
By substantially enlarging the class of input states and linear interferometers for which anomalous bunching can be excluded, we identify a broad regime in which multimode bunching behaves in accordance with physical intuition. At the same time, our analysis demonstrates that anomalous boson bunching is not restricted to polarization-based implementations. In particular, we display a setup with 16 photons in an 18-mode interferometer in which temporal mismatch—an experimentally relevant source of distinguishability—leads to enhanced bunching probabilities compared to the fully indistinguishable scenario.

Taken together, these results sharpen the boundary between interferometric configurations that guarantee the validity of the multimode bunching criterion and those for which it may fail. Our findings, therefore, provide important guidance for the practical use of boson bunching as a validation strategy in boson sampling experiments. More generally, they contribute to a deeper understanding of how particle distinguishability and interferometric structure interplay in multiphoton interference, paving the way for the development of more robust and reliable certification methods for near-term photonic quantum devices.

Despite these advances, several important questions remain open. A natural direction for future research is to determine whether it is possible to rigorously rule out anomalous bunching in other simple configurations. For instance, can anomalous bunching occur in systems with only four, five, or six photons?
Can it arise with only three non-orthogonal sets of fully indistinguishable photons? Another avenue concerns the temporal realization: while our proposed scheme requires 16 photons, it remains an open problem whether anomalous bunching induced by time delays can be observed in configurations involving fewer photons, thereby improving the feasibility of its experimental demonstration.

\section*{Numerical simulations}
\label{sec:code_availability}
The numerical simulations were performed using the package \textsc{Permanents.jl} \cite{seron2023permanentsjl,seron2022bosonsampling}.

\section*{Acknowledgments}
L.P. is a FRIA grantee of the
Fonds de la Recherche Scientifique -- FNRS. N.J.C. acknowledges support from the Fonds de la Recherche Scientifique -- FNRS under project CHEQS within the Excellence of Science (EOS) program. L.N. acknowledges support from FCT-Fundação para a Ciência e a Tecnologia (Portugal) via Project No. CEECINST/00062/2018 and from Horizon Europe project EPIQUE (Grant No. 101135288).

\newpage

\bibliographystyle{unsrt}
\bibliography{main}

@article{MenssenDistinguishability,
  title = {Distinguishability and Many-Particle Interference},
  author = {Menssen, A. J. and Jones, A. E. and Metcalf, B. J. and Tichy, M. C. and Barz, S. and Kolthammer, W. S. and Walmsley, I. A.},
  journal = {Physical Review Letters},
  volume = {118},
  issue = {15},
  pages = {153603},
  numpages = {6},
  year = {2017},
  publisher = {American Physical Society},
  doi = {10.1103/PhysRevLett.118.153603},
  url = {https://link.aps.org/doi/10.1103/PhysRevLett.118.153603}
}

@article{geller2025,
      title={Measuring Multiparticle Indistinguishability with the Generalized Bunching Probability}, 
      author={S. Geller and E. Knill},
      journal={arXiv:2509.04550},
      year={2025},
      eprint={2509.04550},
      archivePrefix={arXiv},
      primaryClass={quant-ph},
      url={https://arxiv.org/abs/2509.04550}, 
}

@article{Math_paper,
title = {A logical implication between two conjectures on matrix permanents},
journal = {Linear Algebra and its Applications},
volume = {725},
pages = {309-318},
year = {2025},
issn = {0024-3795},
url = {https://www.sciencedirect.com/science/article/pii/S0024379525002952},
author = {L. Pioge and K. K. Pietrasz and B. Seron and L. Novo and N. J. Cerf},
}

@article{zhang2013,
author = {Zhang, F.}, 
title = {An analytic approach to a permanent conjecture},
journal = {Linear Algebra and its Applications},
volume = {438},
number = {4},
pages = {1570-1579},
year = {2013},
note = {16th ILAS Conference Proceedings, Pisa 2010},
issn = {0024-3795},
doi = {https://doi.org/10.1016/j.laa.2011.09.034},
}

@article{Marcus1963ThePA,
  title={The permanent analogue of the Hadamard determinant theorem},
  author={M. Marcus},
  journal={Bulletin of the American Mathematical Society},
  year={1963},
  volume={69},
  pages={494-496},
  url={https://api.semanticscholar.org/CorpusID:122357747}
}

@article{pioge2023anomalous,
  title={Anomalous bunching of nearly indistinguishable bosons},
  author={Pioge, L. and Seron, B. and Novo, L. and Cerf, N. J.},
  journal       = {arXiv:2308.12226},
  year          = {2023},
  eprint        = {2308.12226},
  archivePrefix = {arXiv},
  primaryClass  = {quant-ph}
}

@article{zhong2020quantum,
  title={Quantum computational advantage using photons},
  author={Zhong, H.-S. and Wang, H. and Deng, Y.-H. and Chen, M.-C. and others},
  journal={Science},
  volume={370},
  number={6523},
  pages={1460--1463},
  year={2020},
  publisher={American Association for the Advancement of Science}
}

@inproceedings{aaronson2011computational,
  title={The computational complexity of linear optics},
  author={Aaronson, S. and Arkhipov, A.},
  booktitle={Proceedings of the forty-third annual ACM symposium on Theory of computing},
  pages={333--342},
  year={2011}
}

@book{Minc,
  author    = {H. Minc},
  title     = {Permanents},
  publisher = {Cambridge University Press},
  address   = {Cambridge, UK},
  year      = {1978}
}

@article{drury2018,
	title = {A counterexample to a question of {Bapat} \& {Sunder}},
	issn = {1331-4343},
	url = {http://mia.ele-math.com/21-37},
	doi = {10.7153/mia-2018-21-37},
	language = {en},
    volume= {21},
	number = {2},
	urldate = {2023-07-11},
	journal = {Mathematical Inequalities \& Applications},
	author = {Drury, S.},
	year = {2018},
	pages = {517--520},
}

@article{drury2016,
  title={{A counterexample to a question of Bapat and Sunder}},
  author={Drury, S.},
  journal={The Electronic Journal of Linear Algebra},
  volume={31},
  pages={69--70},
  year={2016}
}

@phdthesis{POT,
  title={Matrix functions and the Laplace expansion theorem},
  author={Soules, G. W.},
  year={1966},
  school={University of California, Santa Barbara}
}

@article{chollet,
  title={Is there a permanental analogue to Oppenheim's inequality?},
  author={Chollet, J.},
  journal={The American Mathematical Monthly},
  volume={89},
  number={1},
  pages={57--58},
  year={1982},
  publisher={Taylor \& Francis}
}

@article{lieb1966,
  title={Proofs of some Conjectures on Permanents},
  author={Lieb, E. H.},
  journal={Journal of Mathematics and Mechanics},
  volume={16},
  number={2},
  pages={127--134},
  year={1966},
  publisher={JSTOR}
}

@article{bapat1985majorization,
  title={On majorization and Schur products},
  author={Bapat, R. B. and Sunder, V. S.},
  journal={Linear Algebra and its Applications},
  volume={72},
  pages={107--117},
  year={1985},
  publisher={Elsevier}
}

@article{bapat1986extremal,
  title={An extremal property of the permanent and the determinant},
  author={Bapat, R. B. and Sunder, V. S.},
  journal={Linear Algebra and its Applications},
  volume={76},
  pages={153--163},
  year={1986},
  publisher={Elsevier}
}

@article{bosonbunching,
	doi = {10.1038/s41566-023-01213-0},
	url = {https://doi.org/10.1038%2Fs41566-023-01213-0},
	year = 2023,
	publisher = {Springer Science and Business Media {LLC}},
	volume = {17},
	number = {8},
	pages = {702--709},
	author = {B. Seron and L. Novo and N. J. Cerf},
	title = {Boson bunching is not maximized by indistinguishable particles},
	journal = {Nature Photonics}
}

@article{shchesnovich2016universality,
  title = {Universality of Generalized Bunching and Efficient Assessment of Boson Sampling},
  author = {Shchesnovich, V. S.},
  journal = {Physical Review Letters},
  volume = {116},
  issue = {12},
  pages = {123601},
  numpages = {5},
  year = {2016},
  publisher = {American Physical Society},
  doi = {10.1103/PhysRevLett.116.123601},
  url = {https://link.aps.org/doi/10.1103/PhysRevLett.116.123601}
}

@article{tichy2015_partial_distinguishability,
  title={Sampling of partially distinguishable bosons and the relation to the multidimensional permanent},
  author={Tichy, M. C.},
  journal={Physical Review A},
  volume={91},
  number={2},
  pages={022316},
  year={2015},
  publisher={APS}
}

@article{Zhong_2021,
   title={Phase-programmable gaussian boson sampling using stimulated squeezed light},
   volume={127},
   ISSN={1079-7114},
   url={http://dx.doi.org/10.1103/PhysRevLett.127.180502},
   DOI={10.1103/physrevlett.127.180502},
   number={18},
   journal={Physical Review Letters},
   publisher={American Physical Society (APS)},
   author={Zhong, Han-Sen and Deng, Yu-Hao and Qin, Jian and Wang, Hui and others},
   year={2021},
   month=oct }

@article{Madsen:2022uqz,
    author = "L. S. Madsen and F. Laudenbach and  
 M. F. Askarani and F. Rortais  and others",
    title = "{Quantum computational advantage with a programmable photonic processor}",
    doi = "10.1038/s41586-022-04725-x",
    journal = "Nature",
    volume = "606",
    number = "7912",
    pages = "75--81",
    year = "2022"
}

@article{deng2023gaussianbosonsamplingpseudophotonnumber,
  title = {Gaussian boson sampling with pseudo-photon-number-resolving detectors and quantum computational advantage},
  author={Yu-Hao Deng and Yi-Chao Gu and Hua-Liang Liu and Si-Qiu Gong and others},
  journal = {Phys. Rev. Lett.},
  volume = {131},
  issue = {15},
  pages = {150601},
  numpages = {7},
  year = {2023},
  month = {Oct},
  publisher = {American Physical Society},
  doi = {10.1103/PhysRevLett.131.150601},
  url = {https://link.aps.org/doi/10.1103/PhysRevLett.131.150601}
}

@article{seron2022bosonsampling,
   title={BosonSampling.jl: A Julia package for quantum multi-photon interferometry},
   volume={8},
   ISSN={2521-327X},
   url={http://dx.doi.org/10.22331/q-2024-06-18-1378},
   DOI={10.22331/q-2024-06-18-1378},
   journal={Quantum},
   publisher={Verein zur Forderung des Open Access Publizierens in den Quantenwissenschaften},
   author={Seron, B. and Restivo, A.},
   year={2024},
   pages={1378} 
}

@article{Reck1994,
  title = {Experimental realization of any discrete unitary operator},
  author = {Reck, M. and Zeilinger, A. and Bernstein, H. J. and Bertani, P.},
  journal = {Physical Review Letters},
  volume = {73},
  issue = {1},
  pages = {58--61},
  numpages = {0},
  year = {1994},
  publisher = {American Physical Society},
  doi = {10.1103/PhysRevLett.73.58},
  url = {https://link.aps.org/doi/10.1103/PhysRevLett.73.58}
}

@article{shchesnovich2015tight,
  title={Tight bound on the trace distance between a realistic device with partially indistinguishable bosons and the ideal BosonSampling},
  author={Shchesnovich, V. S.},
  journal={Physical Review A},
  volume={91},
  number={6},
  pages={063842},
  year={2015},
  publisher={APS}
}

@article{young2023atomic,
    title={An atomic boson sampler},
   volume={629},
   ISSN={1476-4687},
   url={http://dx.doi.org/10.1038/s41586-024-07304-4},
   DOI={10.1038/s41586-024-07304-4},
   number={8011},
   journal={Nature},
   publisher={Springer Science and Business Media LLC},
   author={A. W. Young and S. Geller and W. J. Eckner and N. Schine and S. Glancy and E. Knill and A. M. Kaufman},
   year={2024},
  pages={311--316}
}

@article{Seron2024efficientvalidation,
  doi = {10.22331/q-2024-09-19-1479},
  url = {https://doi.org/10.22331/q-2024-09-19-1479},
  title = {Efficient validation of {B}oson {S}ampling from binned photon-number distributions},
  author = {Seron, B. and Novo, L. and Arkhipov, A. and Cerf, N. J.},
  journal = {{Quantum}},
  issn = {2521-327X},
  publisher = {{Verein zur F{\"{o}}rderung des Open Access Publizierens in den Quantenwissenschaften}},
  volume = {8},
  pages = {1479},
  year = {2024}
}

@article{zhang1989notes,
  title={Notes on Hadamard products of matrices},
  author={Zhang, F.},
  journal={Linear and Multilinear Algebra},
  volume={25},
  number={3},
  pages={237--242},
  year={1989},
  publisher={Taylor \& Francis}
}

@article{shchesnovich2016permanent,
  title={The permanent-on-top conjecture is false},
  author={Shchesnovich, V. S.},
  journal={Linear Algebra and its Applications},
  volume={490},
  pages={196--201},
  year={2016},
  publisher={Elsevier}
}

@article{robbio2024centrall,
      title={A central limit theorem for partially distinguishable bosons}, 
      author={M. Robbio and M. G. Jabbour and L. Novo and N. J. Cerf},
      year={2024},
      journal = {arXiv:2404.11518},
      eprint={2404.11518},
      archivePrefix={arXiv},
      primaryClass={quant-ph},
      url={https://arxiv.org/abs/2404.11518}, 
}

@article{hoven2025,
      title={Efficient classical algorithm for simulating boson sampling with heterogeneous partial distinguishability}, 
      author={S. N. van den Hoven and E. Kanis and J. J. Renema},
      year={2025},
      journal={arXiv:2406.17682},
      eprint={2406.17682},
      archivePrefix={arXiv},
      primaryClass={quant-ph},
      url={https://arxiv.org/abs/2406.17682}, 
}

@article{Moylett_2020,
doi = {10.1088/2058-9565/ab5555},
url = {https://doi.org/10.1088/2058-9565/ab5555},
year = {2019},
month = {nov},
publisher = {IOP Publishing},
volume = {5},
number = {1},
pages = {015001},
author = {Moylett, A. E. and García-Patrón, R. and Renema, J. J. and Turner, P. S.},
title = {Classically simulating near-term partially-distinguishable and lossy boson sampling},
journal = {Quantum Science and Technology},
}

@article{renema2019classical,
      title={Classical simulability of noisy boson sampling}, 
      author={J. J. Renema and V. S. Shchesnovich and R. García-Patrón},
      year={2019},
      journal={arXiv:1809.01953},
      archivePrefix={arXiv},
      primaryClass={quant-ph},
      url={https://arxiv.org/abs/1809.01953}, 
}

@article{Qi_2020,
   title={Regimes of Classical Simulability for Noisy Gaussian Boson Sampling},
   volume={124},
   ISSN={1079-7114},
   url={http://dx.doi.org/10.1103/PhysRevLett.124.100502},
   DOI={10.1103/physrevlett.124.100502},
   number={10},
   journal={Physical Review Letters},
   publisher={American Physical Society (APS)},
   author={Qi, Haoyu and Brod, Daniel J. and Quesada, Nicolás and García-Patrón, Raúl},
   year={2020},
   month=mar }

@article{PhysRevA.103.023722,
  title = {Sample-efficient benchmarking of multiphoton interference on a boson sampler in the sparse regime},
  author = {Renema, Jelmer J. and Wang, Hui and Qin, Jian and You, Xiang and Lu, Chaoyang and Pan, Jianwei},
  journal = {Phys. Rev. A},
  volume = {103},
  issue = {2},
  pages = {023722},
  numpages = {7},
  year = {2021},
  month = {Feb},
  publisher = {American Physical Society},
  doi = {10.1103/PhysRevA.103.023722},
  url = {https://link.aps.org/doi/10.1103/PhysRevA.103.023722}
}

@article{Flamini_2020,
doi = {10.1088/2058-9565/aba03a},
url = {https://doi.org/10.1088/2058-9565/aba03a},
year = {2020},
month = {jul},
publisher = {IOP Publishing},
volume = {5},
number = {4},
pages = {045005},
author = {Flamini, Fulvio and Walschaers, Mattia and Spagnolo, Nicolò and Wiebe, Nathan and Buchleitner, Andreas and Sciarrino, Fabio},
title = {Validating multi-photon quantum interference with finite data},
journal = {Quantum Science and Technology},

}

@article{arkhipov2011bosonicbirthdayparadox,
      title={The bosonic birthday paradox}, 
      author={A. Arkhipov and G. Kuperberg},
      year={2011},
      eprint={1106.0849},
      archivePrefix={arXiv},
      primaryClass={quant-ph},
      url={https://arxiv.org/abs/1106.0849}, 
}

@article{PhysRevLett.131.010401,
  title = {Spoofing Cross-Entropy Measure in Boson Sampling},
  author = {Oh, Changhun and Jiang, Liang and Fefferman, Bill},
  journal = {Phys. Rev. Lett.},
  volume = {131},
  issue = {1},
  pages = {010401},
  numpages = {7},
  year = {2023},
  month = {Jul},
  publisher = {American Physical Society},
  doi = {10.1103/PhysRevLett.131.010401},
  url = {https://link.aps.org/doi/10.1103/PhysRevLett.131.010401}
}

@article{PRXQuantum.5.040312,
  title = {Linear Cross-Entropy Certification of Quantum Computational Advantage in Gaussian Boson Sampling},
  author = {Mart\'{\i}nez-Cifuentes, Javier and de Guise, Hubert and Quesada, Nicol\'as},
  journal = {PRX Quantum},
  volume = {5},
  issue = {4},
  pages = {040312},
  numpages = {44},
  year = {2024},
  month = {Oct},
  publisher = {American Physical Society},
  doi = {10.1103/PRXQuantum.5.040312},
  url = {https://link.aps.org/doi/10.1103/PRXQuantum.5.040312}
}

@article{doi:10.1142/S021974991560028X,
author = {Bentivegna, Marco and Spagnolo, Nicol\`{o} and Vitelli, Chiara and Brod, Daniel J. and others},
title = {Bayesian approach to Boson sampling validation},
journal = {International Journal of Quantum Information},
volume = {12},
number = {07n08},
pages = {1560028},
year = {2014},
doi = {10.1142/S021974991560028X},

eprint = { 
    
        https://doi.org/10.1142/S021974991560028X
    
    

}}

@article{Go_2025,
   title={Quantum Computational Advantage of Noisy Boson Sampling with Partially Distinguishable Photons},
   volume={6},
   ISSN={2691-3399},
   url={http://dx.doi.org/10.1103/rflv-gc66},
   DOI={10.1103/rflv-gc66},
   number={3},
   journal={PRX Quantum},
   publisher={American Physical Society (APS)},
   author={Go, Byeongseon and Oh, Changhun and Jeong, Hyunseok},
   year={2025},
   month=sep }

@article{PhysRevLett.120.220502,
  title = {Efficient Classical Algorithm for Boson Sampling with Partially Distinguishable Photons},
  author = {Renema, J. J. and Menssen, A. and Clements, W. R. and Triginer, G. and Kolthammer, W. S. and Walmsley, I. A.},
  journal = {Phys. Rev. Lett.},
  volume = {120},
  issue = {22},
  pages = {220502},
  numpages = {5},
  year = {2018},
  publisher = {American Physical Society},
  doi = {10.1103/PhysRevLett.120.220502},
  url = {https://link.aps.org/doi/10.1103/PhysRevLett.120.220502}
}

@article{anguita2025experimentalvalidationbosonsampling,
doi = {10.1088/2058-9565/adecbc},
url = {https://doi.org/10.1088/2058-9565/adecbc},
year = {2025},
publisher = {IOP Publishing},
volume = {10},
number = {3},
pages = {035062},
author={M. C. Anguita and A. Camillini and S. Marzban and M. Robbio and B. Seron and L. Novo and J. J. Renema},
title = {Experimental validation of boson sampling using detector binning},
journal = {Quantum Science and Technology},
}

@article{bressanini2024gaussianbosonsamplingvalidation,
  title = {Binned-detector probability distributions for Gaussian boson sampling validation},
  author={G. Bressanini and B. Seron and L. Novo and N. J. Cerf and M. S. Kim},
  journal = {Physical Review A},
  volume = {112},
  issue = {1},
  pages = {012610},
  numpages = {14},
  year = {2025},
  publisher = {American Physical Society},
    doi = {10.1103/PhysRevA.112.012610},
  url = {https://link.aps.org/doi/10.1103/jqvf-pm1r}
}

@article{seron2023permanentsjl,
    author = {Seron, B.},
    title = {Permanents.jl: A {J}ulia package for computing permanents},
    year = {2023},
    note = {Available at \href{https://github.com/benoitseron/Permanents.jl}{\textsc{Permanents.jl}}}, 
}

@article{Pate01011996,
author = {T. H. Pate},
title = {Group algebras, monotonicity, and the lieb permanent inequality},
journal = {Linear and Multilinear Algebra},
volume = {40},
number = {3},
pages = {207--220},
year = {1996},
publisher = {Taylor \& Francis},
doi = {10.1080/03081089608818438},
}

@article{oppenheim1930inequalities,
  title={Inequalities connected with definite Hermitian forms},
  author={Oppenheim, A.},
  journal={Journal of the London Mathematical Society},
  volume={1},
  number={2},
  pages={114--119},
  year={1930},
  publisher={Wiley Online Library}
}

@article{drury2017real,
  title={A real counterexample to two inequalities involving permanents},
  author={Drury, S.},
  journal={Mathematical Inequalities \& Applications},
  volume    = {20},
  number    = {2},
  pages     = {349--352},
  year      = {2017},
  doi       = {10.7153/mia-20-23},
}

\onecolumngrid

\appendix

\section{Derivation of the multimode bunching probability}
\label{appendix:Derivation_of_the_Multimode_bunching_probability}

The multimode bunching probability is given by Eq.~\eqref{eq:multimode_boson_bunching}, as shown in Ref.~\cite{shchesnovich2016universality}. In this appendix, we present an alternative -- very simple -- derivation of Eq.~\eqref{eq:multimode_boson_bunching} in the framework of first quantization. Prior to symmetrization, the $n$-photon input state is given by 
$\ket{\psi^{in}}=\bigotimes^{n}_{i=1}\ket{i,\phi_i}$, $\ket{i}$ corresponds to the spatial mode of the $i$th photon, while $\ket{\phi_i}$ correspond to its internal modes.
 After symmetrization, the properly symmetrized state reads 
\begin{equation}
\label{eq:input_state}
     \ket{\psi^{in}}= \frac{1}{\sqrt{n!}}\sum_{\sigma \in S_{n}}\Big(\bigotimes^{n}_{i=1}\ket{\sigma(i),\phi_{\sigma(i)}}\Big).
\end{equation}
In this context, the factor $\sqrt{n!}$ provides the normalization, while $ \sigma $ denotes an element of the symmetric group $S_n$. The action of the linear interferometer is represented by a unitary matrix $U$ with corresponding single-photon operator $\hat{u}$ and $n$-photon operator $\hat{U}=\hat{u}^{\otimes n}$, so that the output state is $\hat{U}\ket{\psi^{in}}$, where
\begin{equation}
\hat{u}\ket{i,\phi_i}=\sum_{k=1}^{m} U_{k,i}\ket{k,\phi_i} ,~\forall i.
\end{equation}
$U_{k,i}$ is the probability amplitude for a single photon entering the $i$th spatial mode to exit in the $k$th spatial mode. The probability that all photons bunch into a subset of output modes $\kappa$ is obtained by projecting the output state onto the subspace where photons occupy only modes in $\kappa$
\begin{equation}
P_\kappa=\bra{\psi^{in}}\hat{U}^\dagger\hat{\Pi}_{\kappa}\hat{U}\ket{\psi^{in}},
\end{equation}
where $\hat{\Pi}_{\kappa}$ is the $n$-photon projector onto the subspace in which each photon occupies an output mode in $\kappa$. We next define the single-photon operator $\hat{H}$ acting solely on spatial modes such that $\hat{U}^\dagger\hat{\Pi}_{\kappa}\hat{U}= (\hat{H}\otimes \hat{\mathbb{1}})^{\otimes n}$, where $\hat{\mathbb{1}}$ denotes the identity operator on the internal degree of freedom. Concretely, $\hat{H}$ corresponds to the $n\times n$ matrix,
\begin{equation}
H_{i,j}=\sum_{k\in\kappa}U^{*}_{k,i}U_{k,j},~~\forall i,j.
\end{equation}
The bunching probability can then be expressed as
\begin{equation}
    P_\kappa=\frac{1}{n!}\sum_{\sigma,\epsilon \in S_{n}}\prod_{i=1}^n\bra{\sigma(i)}\hat{H}\ket{\epsilon(i)}\braket{\phi_{\sigma (i)}}{\phi_{\epsilon(i)}}.
\end{equation}
We now introduce the \textit{distinguishability matrix}, denoted $S$, which encodes all the overlaps of the photons' internal states  $S_{i,j}=\braket{\phi_i}{\phi_j}$. Using these new notations, the bunching probability can be rewritten as
\begin{equation}
    P_\kappa=\frac{1}{n!}\sum_{\sigma,\epsilon \in S_{n}}\prod_{i=1}^n H_{\sigma(i),\epsilon(i)}S_{\sigma(i),\epsilon(i)}.
\end{equation}
Using the definition of the matrix permanent and expanding one sum after the other, we finally obtain 
\begin{equation}
    P_\kappa= \perm{H\odot S},
\end{equation}
the symbol "$\odot$" denotes the Hadamard product of two matrices, defined as $(H\odot S)_{i,j}=H_{i,j}S_{i,j}$. This last equation is the standard equation for the multimode bunching initially introduced in Ref.~\cite{shchesnovich2016universality}, and it reveals the fundamental connection between matrix permanents and quantum bunching phenomena.

\section{Validity regimes for the Bapat-Sunder inequality}
\label{appendix:derivative}
To support the analysis of Conjecture~\ref{conjm:bapatSunder_1}, we develop a general differentiation formula for the permanent of a $p.s.d.h.$ matrix, i.e., for $\derivate{}{x}\perm{G(x)}$.
Let $G(x)$ be an $n\times n$ $p.s.d.h.$ matrix depending on a single parameter $x$. Here, we present the general derivation, while in subsequent subsections, we instantiate it for physically relevant distinguishability scenarios. We start by applying the Laplace expansion to the first row of 
$G(x)$ \cite{Minc}: 
\begin{align}
    \derivate{\perm{G(x)}}{x}&=\derivate{}{x}\sum_{j=1}^n G_{1,j}\perm{G(1;j)}\\
    &=\sum_{j=1}^n\derivate{G_{1,j}}{x}\perm{G(1;j)}+\sum_{j=1}^n G_{1,j}\derivate{}{x}\perm{G(1;j)}.
\end{align}
In general, $G(a;b)$ denotes the $(n - 1)\times (n - 1)$ submatrix of $G$ obtained by deleting the $a$th row and $b$th column of $G$. To compute the second term, we apply the Laplace expansion once more to the first row of $G(1;j)$, i.e., to the second row of $G$,
\begin{align}
    \sum_{j=1}^n G_{1,j}\derivate{\perm{G(1;j)}}{x}&=\sum_{j= 1}^nG_{1,j}\derivate{ }{x} 
    \sum_{k\neq j}^n G_{2,k}\perm{G(1,2;j,k)}\\
    &=\sum_{j= 1}^nG_{1,j}\sum_{k \neq j}^n \derivate{G_{2,k}}{x}\perm{G(1,2;j,k)}+ \sum_{j= 1}^nG_{1,j} 
    \sum_{k\neq j}^n G_{2,k}\derivate{ }{x}\perm{G(1,2;j,k)}\\
    &=\sum_{k= 1}^n \derivate{G_{2,k}}{x}\sum_{j \neq k}^n G_{1,j} \perm{G(1,2;j,k)}+ \sum_{j= 1}^n \sum_{k\neq j}^n G_{1,j} 
     G_{2,k}\derivate{ }{x}\perm{G(1,2;j,k)}\\
     &=\sum_{k= 1}^n \derivate{G_{2,k}}{x}\perm{G(2;k)}+ \sum_{j= 1}^n \sum_{k\neq j}^n G_{1,j} 
     G_{2,k}\derivate{ }{x}\perm{G(1,2;j,k)}.
\end{align}
The Laplace expansion is invoked again in the last step, thus reconstructing $G(2;k)$ starting from $\sum_j G_{1,j}\perm{G(1,2;j,k)}$. By iterating this procedure until all rows and columns of the right term are exhausted, and retaining at each stage the first sum, we finally obtain
\begin{equation}
\label{eq:derivative}
    \derivate{\perm{G(x)}}{x}=\sum_{i,j}\derivate{G_{i,j}}{x}\perm{G(i;j)}.
\end{equation}

\subsection{$x_i$-model of distinguishability}
\label{appendix:x_i-model}
Let us consider the first natural generalization of the $x$-model. In this configuration, the photons are initially in the same internal state $\ket{\chi}$ and are interpolated toward fully distinguishable states
$\{\ket{\eta_1}\dots\ket{\eta_n}\},$ with $\braket{\eta_i}{\chi}=0$ and $\braket{\eta_i}{\eta_j}=\delta_{i,j},~\forall i,j$. Here, we allow the interpolation coefficient to vary across photons, which offers a more flexible description of partial distinguishability. This yields the following states and distinguishability matrix:
\begin{align}
\label{eq:xi_model_appendix}
    \ket{\phi_i}&=x_i\ket{\chi}+\sqrt{1-|x_i|^2} \ket{\eta_i}\\
    S^{\bm{x}}_{i,j}&=x_i^{*}x_j(1-\delta_{i,j})+\delta_{i,j}.
\end{align}
By absorbing the phases into the states 
$\ket{\eta_i}$, we may, without loss of generality, restrict to $x_i\in[0,1]$, for all $i$. To prove that $S^{\bm{x}}$ satisfies Conjecture~M\ref{conjm:bapatSunder_1}, we need to show that, for any $p.s.d.h.$ matrix $H$, we have
\begin{equation}
    \perm{H\odot S^{\bm{x}}}\leq \perm{H}. 
\end{equation}
To this end, we will show that the left-hand side of the previous equation is maximized when $x_i=1,~\forall i$, i.e., when $S^{\bm{x}}=\mathbb{E}$. We therefore compute the derivative of the bunching probability with respect to $x_i$, and show that it is nonnegative. First, note that
\begin{equation}
    \perm{H\odot S^{\bm{x}}}=\prod_{k=1}^{n}x_k^2~\perm{G},
\end{equation}
where $G_{i,j}=H_{i,j}\big(\frac{1}{x_{i}^{2}}\delta_{i,j}+(1-\delta_{i,j})\big)$. To obtain this last equation, we used the fact that a common scalar factor in any row or column can be factored out of the permanent. By differentiating, we get
\begin{equation}
    \frac{\partial \perm{H\odot S^{\bm{x}}}}{\partial x_{i}}=2x_i \prod_{k\neq i }^n x_{k}^{2}~\perm{G}+ \prod_{k= 1 }^n x_{k}^{2}\frac{\partial \perm{G}}{\partial x_i}.
\end{equation}

According to Eq.~\eqref{eq:derivative}, the derivative reads
\begin{equation}
     \derivate{\perm{G}}{x_i}=-\frac{2}{x_{i}^{3}}H_{i,i}~\perm{G(i;i)}.
\end{equation}

By using Lieb’s inequality, i.e., $\perm{H}\geq H_{i,i}\perm{H(i;i)}$ \cite{lieb1966}, we obtain
\begin{align}
    \derivate{\perm{H\odot S^{\bm{x}}}}{x_i}&=2x_i \perm{G}\prod_{k\neq i }^n x_{k}^{2} -\frac{2H_{i,i}}{x_{i}^{3}}\perm{G(i;i)}\prod_{k= 1 }^n x_{k}^{2}\\
    &=\frac{2}{x_i}\Big(\perm{H}-H_{i,i}~\perm{H(i;i)}\Big)\\
    &\geq 0.
\end{align}

Since $x_i \in [0,1]$, and $\derivate{}{x_i}\perm{H\odot S^{\bm{x}}}\geq 0,~\forall i$, the bunching probability is maximized for $x_i=1$, $\forall i$, i.e. $S^{\bm{x}}=\mathbb{E}$.\qed 

\subsection{Two non-orthogonal sets of fully indistinguishable particles}
\label{appendix:2_set}

We now turn to the physically relevant configuration of two groups of partially distinguishable bosons. The first $k$ particles are in the state $\ket{\chi}$, while the remaining $(n-k)$ particles are in the state $\ket{\eta}$, with $\braket{\chi}{\eta}=x$. Without loss of generality, we may restrict to $x\in[0,1]$. The corresponding distinguishability matrix has the following block structure:
\begin{equation}
    S=\begin{pmatrix}
        \mathbb{E}^{k} & X \\
        X^{\dagger} & \mathbb{E}^{(n-k)}
    \end{pmatrix},
\end{equation}
Here, $\mathbb{E}^{a}$ is the $a\times a$ matrix filled with ones, and $X$ is the $k\times(n-k)$ matrix with entries $X_{i,j}=x$ for all $i,j$. As in the previous subsection, to prove that $S$ satisfies Conjecture~M\ref{conjm:bapatSunder_1}, we show that for any $n\times n$ $p.s.d.h.$ matrix $H$,
\begin{equation}
\derivate{\perm{H\odot S(x)}}{x}\geq 0,~\forall x\in[0,1]. 
\end{equation} 
For brevity, set $G(x)=S^A\odot S(x)$. Using Eq.~\eqref{eq:derivative}, the derivative becomes
\begin{align}
  \derivate{\perm{G(x)}}{x}&=\sum_{i,j=1}^n\derivate{G_{i,j}}{x}\perm{G(i;j)}\\
  &=\sum_{i=1}^k \sum_{j=k+1}^n H_{i,j}\perm{G(i;j)} +\sum_{i=k+1}^n \sum_{j=1}^k H_{i,j}\perm{G(i;j)}\\
  &=\frac{1}{x}\bigg(\sum_{i=1}^k \sum_{j=k+1}^nG_{i,j}\perm{G(i;j)} +\sum_{i=k+1}^n \sum_{j=1}^k G_{i,j}\perm{G(i;j)}\bigg).
\end{align}
Applying the Laplace expansion, the last expression can be rewritten as
\begin{align}
\label{eq:before_pate}
    \derivate{\perm{G(x)}}{x}&=\frac{1}{x}\bigg( k~\perm{G}-\sum_{a=1}^k \sum_{b=1}^kG_{i,j}\perm{G(i;j)} \bigg)\nonumber \\
    &+\frac{1}{x}\bigg( (n-k)~\perm{G}-\sum_{i=k+1}^n \sum_{j=k+1}^n G_{i,j}\perm{G(i;j)}\bigg).
\end{align}
By Theorem 2 from Ref.~\cite{Pate01011996}, both terms in Eq.~\eqref{eq:before_pate} are nonnegative,
and hence
\begin{equation}
\derivate{\perm{H\odot S(x)}}{x}\geq 0,~\forall x\in[0,1]. \qed
\end{equation}

\section{Time-delay-induced distinguishability}
\label{appendix:Time_delay}
In this appendix, we consider time delays between photon arrival times. This phenomenon is a dominant source of distinguishability in current photonic platforms.
A standard model for multiphoton interference assumes that each photon is described by a Gaussian wave packet centered at times $\{t_1,\ldots,t_n\}$, forming a real vector $\bm{t}$, where $\bm{\tau}=\bm{t}/||\bm{t}||$ denotes its normalized version. All wave packets are assumed to have the same variance, $\sigma^2$. Defining $d=||\bm{t}||/(2\sigma)$, the distinguishability matrix for this model reads \cite{MenssenDistinguishability}:
\begin{equation}
    S^{\bm{\tau}}_{i,j}=e^{-\frac{{(t_i-t_j)^2}}{4\sigma^2}}=e^{-(\tau_i-\tau_j)^2 d^2},~\forall i,j.
\end{equation}
To simplify notation, define $G(d)=H\odot S^{\bm{\tau}}(d)$. The objective of this appendix is to compute $\derivate{}{d}\perm{G(d)}$ and 
$\left .\frac{\partial^2}{\partial d^2} \perm{G(d)}\right|_{d=0} $, in order to establish the occurrence of anomalous bunching in this configuration (see Sec.~\ref{sec:Time_delay_distinguishability}). As a first step toward computing these derivatives, note that, for each $i,j$,
\begin{equation}
\derivate{G_{i,j}(d)}{d}=-2(\tau_i-\tau_j)^2 d~G_{i,j}(d).    
\end{equation} 
Using this identity together with Eq.~\eqref{eq:derivative}, the derivative of the bunching probability is
\begin{align}
    \derivate{}{d}\perm{G(d)}
    &= \sum_{i=1}^n \sum_{j=1}^n-2(\tau_i-\tau_j)^2d~G_{i,j}{\perm{G(i;j)}}\\ 
    &=2d\sum_{i=1}^n \sum_{j=1}^n (2\tau_i F^G_{i,j}\tau_j - \tau_i^2F^{G}_{i,j}-\tau_j^2F^{G}_{i,j}).
\end{align}
Here, $F^{G}$ denotes the $n\times n $ $p.s.d.h.$ matrix defined by $F_{i,j}^{G}=G_{i,j}\perm{G(i;j)}$, $\forall i,j$. Using the Laplace expansion, $\sum_{i}F^{G}_{i,j}=\sum_{j}F^{G}_{i,j}=\perm{G}$ (for all 
$j$ and $i$, respectively), the derivative simplifies to
\begin{equation}
    \derivate{}{d}\perm{G(d)}
    = 4d\big(\bm{\tau}^T F^{G}\bm{\tau}-\perm{G(d)}\bigr).\\
\end{equation}
Differentiating this expression once more and evaluating at $d=0$ yields
\begin{equation}
     \left .\frac{\partial^2\perm{G(d)}}{\partial d^2} \right|_{d=0} =4\bigl(\bm{\tau}^T F^{G(0)}\bm{\tau}-\perm{G(0)}\bigr),
\end{equation}
This concludes the appendix.
\newpage

\end{document}